\documentclass[journal=ancac3,layout=twocolumn]{achemso}
\usepackage{siunitx}
\usepackage{graphicx}
\usepackage{amsmath}
\usepackage{rotating}
\usepackage{textcomp}
\usepackage{xcolor}
\mciteErrorOnUnknownfalse % used to silence errors of citations in SI

\author{Chak Kui Wong}
\affiliation{Physical and Theoretical Chemistry Laboratory, Department of Chemistry,
University of Oxford, South Parks Road, Oxford, OX1 3QZ, UK}
\author{Chuyan Tang}
\affiliation{Physical and Theoretical Chemistry Laboratory, Department of Chemistry,
University of Oxford, South Parks Road, Oxford, OX1 3QZ, UK}
\author{John S. Schreck}
\affiliation{National Center for Atmospheric Research,
Computational and Information Systems Laboratory,
850 Table Mesa Drive, Boulder, CO 80305, USA}
\author{Jonathan P. K. Doye}
\email{jonathan.doye@chem.ox.ac.uk}
\affiliation{Physical and Theoretical Chemistry Laboratory, Department of Chemistry,
University of Oxford, South Parks Road, Oxford, OX1 3QZ, UK}

\title[DNA origami free-energy landscapes]{Characterizing the free-energy landscapes of DNA origamis}


\begin{document}

\begin{abstract}
We show how coarse-grained modelling combined with umbrella sampling using distance-based order parameters can be applied to compute the free-energy landscapes associated with mechanical deformations of large DNA nanostructures. We illustrate this approach for the strong bending of DNA nanotubes and the potentially bistable landscape of twisted DNA origami sheets. The homogeneous bending of the DNA nanotubes is well described by the worm-like chain model; for more extreme bending the nanotubes reversibly buckle with the bending deformations localized at one or two ``kinks''. For a twisted one-layer DNA origami, the twist is coupled to the bending of the sheet giving rise to a free-energy landscape that has two nearly-degenerate minima that have opposite curvatures. By contrast, for a two-layer origami, the increased stiffness with respect to bending leads to a landscape with a single free-energy minimum that has a saddle-like geometry. The ability to compute such landscapes is likely to be particularly useful for DNA mechanotechnology and for understanding stress accumulation during the self-assembly of origamis into higher-order structures.
\end{abstract}

\section{Introduction}
DNA nanotechnology aims to design and construct self-assembled nanoscale DNA structures through utilizing the specificity of Watson-Crick base-pairing and the structural rigidity of the DNA double helix \cite{Fan17,Seeman17}. Probably the most commonly used method is DNA origami \cite{Rothemund06, Dey21}. In this method, a long viral DNA single strand, often termed the ``scaffold'' strand, is folded into the target structure with the aid of many different ``staple'' strands that are programmed to hybridize to multiple specific domains of the scaffold strand.

Probably the most basic property of a DNA origami is its equilibrium average structure and the thermal fluctuations about that average. The most detailed experimental views of origami structure can be obtained from cryo-EM studies \cite{Bai12, Kube20} with lower-resolution imaging available from TEM and AFM. Alternatively, computer modelling can be used to gain structural insights into DNA nanostructures. All-atom molecular dynamics simulations can provide the highest level of detail \cite{Yoo13,Li15,Gopfrich16,Maffeo16,Slone16,Lee17}, but are computationally very expensive because of the large size of DNA origamis. On the other hand, coarse-grained models such as oxDNA \cite{Ouldridge11,Sulc12,Snodin15}, mrDNA \cite{Maffeo20}, CanDo \cite{Kim12,Castro11} and SNUPI\cite{Lee21} provide a lower level of detail, but are computationally much less expensive and so can be routinely used to gain insight into origami structure \cite{Snodin19,Berengut19}, even for those that are very flexible and diffusively explore their configuration space \cite{Sharma17,Shi17,Zhou18,Huang19,Shi19}. 

The free-energy landscapes associated with more extreme fluctuations are also potentially important. For example, such landscapes can describe the response of DNA origami to stresses, and so are particularly pertinent to the field of DNA mechanotechnology \cite{Blanchard19} that uses DNA nanodevices to sense \cite{Dutta18} or apply forces,\cite{Nickels16b} 
and to origamis in which internal stresses are used to modulate the structure.\cite{Dietz09,Liedl10,Zhou14} 
Furthermore, for origamis with multiple stable states,\cite{Stammers18,Zhou15,Shrestha16} the landscapes would allow the mechanisms of interconversion and the free-energy barriers between the states to be characterized.
Additionally, for origamis that are designed to assemble into higher-order structures, the monomer landscapes could help predict the thermodynamics of stress accumulation, particularly for examples where the stress accumulation is a design feature, for example, to make the assembly self-limiting. \cite{Berengut20,Hagan20}

One of the most common methods for computing free-energy landscapes is umbrella sampling \cite{Torrie77}, in which a bias is applied during the simulations to allow sampling of high free-energy conformations. For example, for small DNA systems it has been used to study processes such as stacking \cite{Maffeo12,Hase16}, condensation \cite{Kang18,Cortini17}, and unwinding \cite{Liebl17} in all-atom detail. Coarse-grained models allow this approach to be applied to larger systems and slower processes. For example, the oxDNA model has been used to study the free-energy landscapes for a wide variety of processes involving assembly and/or mechanical deformation both for basic DNA motifs\cite{Harrison19,Mosayebi15,Romano13} and for small DNA nanodevices. \cite{Ouldridge10,Ouldridge13,Khara18} 

Here, we apply this approach to sample some example free-energy landscapes associated with full-size origamis. Such calculations, although computationally expensive, are feasible both because we are using a coarse-grained model (oxDNA)
and because we are considering the free-energy landscapes for structural variables where equilibration mainly involves relatively fast processes associated with mechanical relaxation rather than slow processes such as hybridization.

The first example we consider is the bending and stretching of DNA nanotubes\cite{Pfeifer18b} under external stress. 
One might expect that the behaviour will be somewhat analogous to that for double-stranded DNA. When weakly bent, duplex DNA can be described with the worm-like chain (WLC) model,\cite{Bustamante94} but under strong bending, its behaviour deviates from the model due to localized buckling to form a kink that typically involves broken base pairs.\cite{Vafabakhsh12b,Mitchell11,Harrison15}
Previous analysis of such nanotubes have shown that the WLC model provides a good description of their thermal fluctuations, \cite{Chhabra20}
but it will be interesting to explore the degree of bending that can be achieved before kink formation, the structure of the kinks that form and whether such buckling leads to irreversible damage to the nanotubes.

In the second example we consider the coupling between twist and curvature in twisted origami sheets.
For an inextensible sheet, twist must lead to curvature, where there are two possible states that differ in the sign of the curvature and the diagonal about which bending occurs. That such single-layer DNA origami sheets exhibit curvature has been confirmed by SAXS\cite{Baker18} and cryoEM\cite{Kube20}; AFM studies have also attempted to identify the preferred direction of curvature. \cite{Marchi14,Stammers18}
Here, we characterize the free-energy landscapes of twisted DNA origami sheets to probe whether there are indeed two free-energy minima; we also explore the effects of sheet thickness on this potential bistability. 

To compute these free-energy landscapes we use oxDNA, which is a coarse-grained model of DNA at the nucleotide level.\cite{Ouldridge11,Sulc12,Snodin15} It has been designed to reproduce the structure, thermodynamics and mechanics of double- and single-stranded DNA. This allows it to capture many biophysical properties of DNA;\cite{Doye13} those particularly relevant to the current study include the response of double-stranded DNA to bending stress\cite{Harrison15,Harrison19} and the fraying or breaking of DNA base-pairs under stress more generally.\cite{Mosayebi15} Moreover, the latest version of the model is able to accurately model the structural properties of DNA origami,\cite{Snodin19} and so has been applied to study a wide range of different types of DNA origami.\cite{Sharma17,Shi17,Benson18,Berengut19,Berengut20} 
Given the above, we are confident in using oxDNA to calculate the free-energy landscapes for our chosen DNA origamis, as detailed in the rest of this article.

\section{Results and Discussion}
\subsection{DNA nanotube mechanics}

\begin{figure*}[ht!]
    \centering
    \includegraphics[width=170mm]{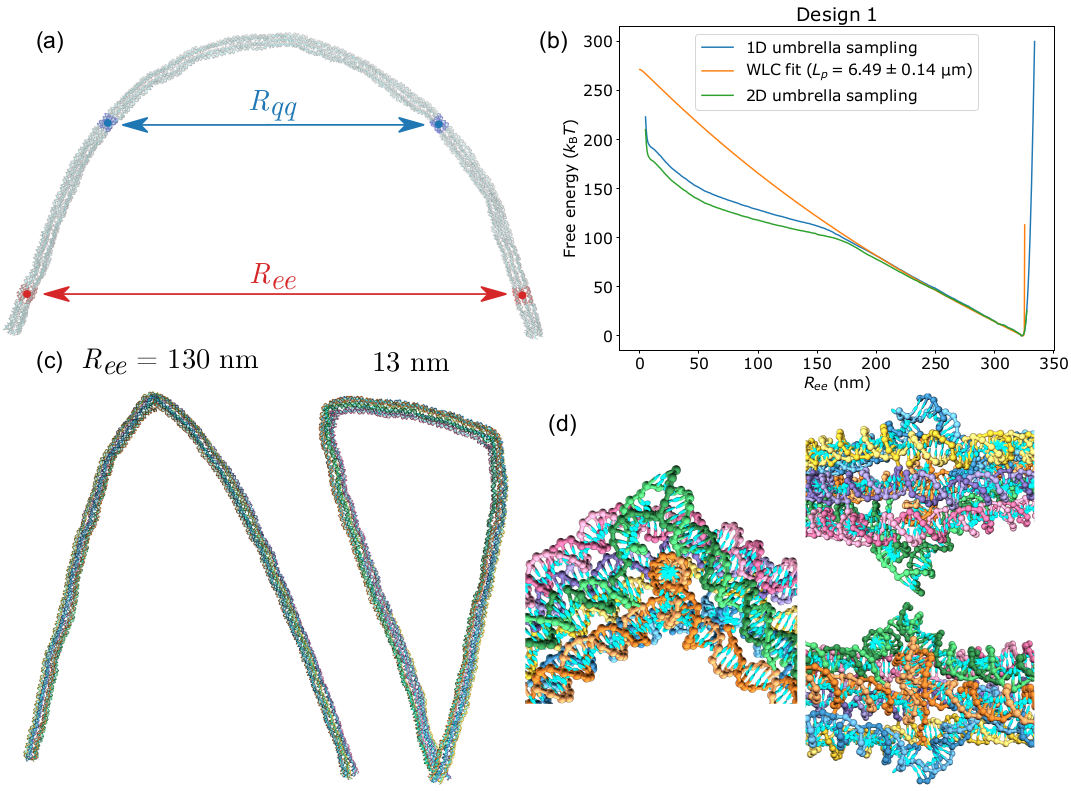}
    \caption{(a) The end-to-end distance $R_{ee}$ and the \textonequarter--\textthreequarters\ distance $R_{qq}$ that are used as order parameters in the umbrella sampling illustrated for the design 1 nanotube with $R_{ee}=210\,$nm. Each distance is defined with respect to the centres of mass of the shaded sets of nucleotides. As the helices at the ends of an origami tend to splay out more, we use sets of nucleotides slightly in from the ends to define $R_{ee}$.
    (b) Free-energy landscape of design 1 as a function of $R_{ee}$ computed
 by 1- and 2-dimensional umbrella sampling.
    The weak-bending regime of the landscape can be fitted to the WLC model (see Table S2 for the fitting parameters). 
    (c) Further representative configurations at different $R_{ee}$. (d) Close-ups of the kink for the $R_{ee}=130$\,nm configuration. Views are from the side, outside and inside of the kink.}
    \label{fig:A(r)}
\end{figure*}

The DNA nanotubes that we study here are all six-helix bundles with a hexagonal cross-section. Six-helix bundles have been particularly well characterized\cite{Schiffels13,Kauert11,Lee19b,Kube20}, and commonly used --- example applications include as chiral liquid-crystal mesogens\cite{Siavashpouri17} and stiff handles for force spectroscopy.\cite{Pfitzner13} For two of the examples studied here, the elastic mechanical properties have been measured experimentally\cite{Schiffels13,Pfitzner13}. First, we consider the general properties of the free-energy landscapes that describe the nanotube bending by considering a single example, before focusing on how the details of the designs can affect these properties.

To describe the degree of bending, we define an end-to-end distance $R_{ee}$ 
for the nanotubes (Fig.\ \ref{fig:A(r)}(a)). This ``order parameter'' is then used to bias the systems 
to sample extremely bent states. The umbrella sampling works by running multiple simulations each of which samples a different $R_{ee}$ ``window'' by adding a harmonic ``umbrella'' potential that constrains $R_{ee}$ to be near to the minimum of that particular potential. A best estimate of the probability distribution $p(R_{ee})$ can then be reconstructed from the distributions in the individual simulations. 
The free-energy landscape is then simply $A(R_{ee})=-k_\mathrm{B} T \log(p(R_{ee}))+c$ where we choose $c$ such that the free-energy global minimum has a value of 0. 

The free-energy landscape that results from this ``1-dimensional'' umbrella sampling is shown in Fig.\ \ref{fig:A(r)}(b) for the six-helix bundle studied in Ref.\ \citenum{Schiffels13} along with some representative configurations at different $R_{ee}$ in Fig.\ \ref{fig:A(r)}(a) and (c). The landscape has a minimum close to the contour length of the nanotube; this minimum is asymmetric as it costs more to extend $R_{ee}$ by stretching the nanotube than to reduce $R_{ee}$ by bending the nanotube. That being said the nanotube is still extremely stiff towards bending; for example, it costs more than $100\,k_\mathrm{B}T$ to reduce $R_{ee}$ by 50\%. The landscape has a clear change in slope at about 160\,nm that corresponds to the onset of ``kink'' formation, where it becomes more favourable to localize much of the stress at a defect rather than the system being homogeneously bent. Finally, there is a sharp rise in the free-energy as $R_{ee}$ approaches zero because the nucleotides at the two ends then begin to overlap.

In the weak to moderate bending regime, we expect the behaviour of the nanotube to be well described by the WLC model. Indeed, the fit to the free-energy landscape in this regime using the WLC formula for $p(R_{ee})$ given in Ref.\ \citenum{Becker10} accurately describes the landscape up until the onset of kink formation. 
Note that this is despite significant local inhomogeneities in the nanotube; for example, bending of the individual helices at nicks and junctions is likely to be somewhat easier, and the relative motion of the helices is much more tightly constrained at junctions. As observed in Ref.\ \citenum{Chhabra20}, although these inhomogeneities lead to significant deviations from worm-like chain behaviour at short length scales, as longer length scales are considered the nanotubes behave more and more like an ideal worm-like chain.
Note also that the WLC form cannot capture the behaviour of the landscape when $R_{ee}$ is larger than that for the free-energy minimum, i.e.\ when the nanotube is stretched; this is simply because the formula is for the standard form of the WLC model which assumes the ``chain'' is inextensible.

In the strong bending regime, a kink develops in the nanotube, which leads to deviation from WLC behaviour. The formation of the kink can be caused by the localization of the bending stress in a small region in order to relieve the bending stress in the rest of nanotube. A close-up of a kink is shown in Fig.\ \ref{fig:A(r)}(d). Helices in the outer part of the kink are stretched, while those in the inner part are compressed and forced out of the helical axis.
At very small $R_{ee}$ it can become favourable for the nanotube to adopt a configuration with two kinks with bend angles of near to 90$^\circ$ rather than a configuration with a single kink with a bend angle near to 180$^\circ$.

\begin{figure}[t]
    \centering
    \includegraphics[width=84mm]{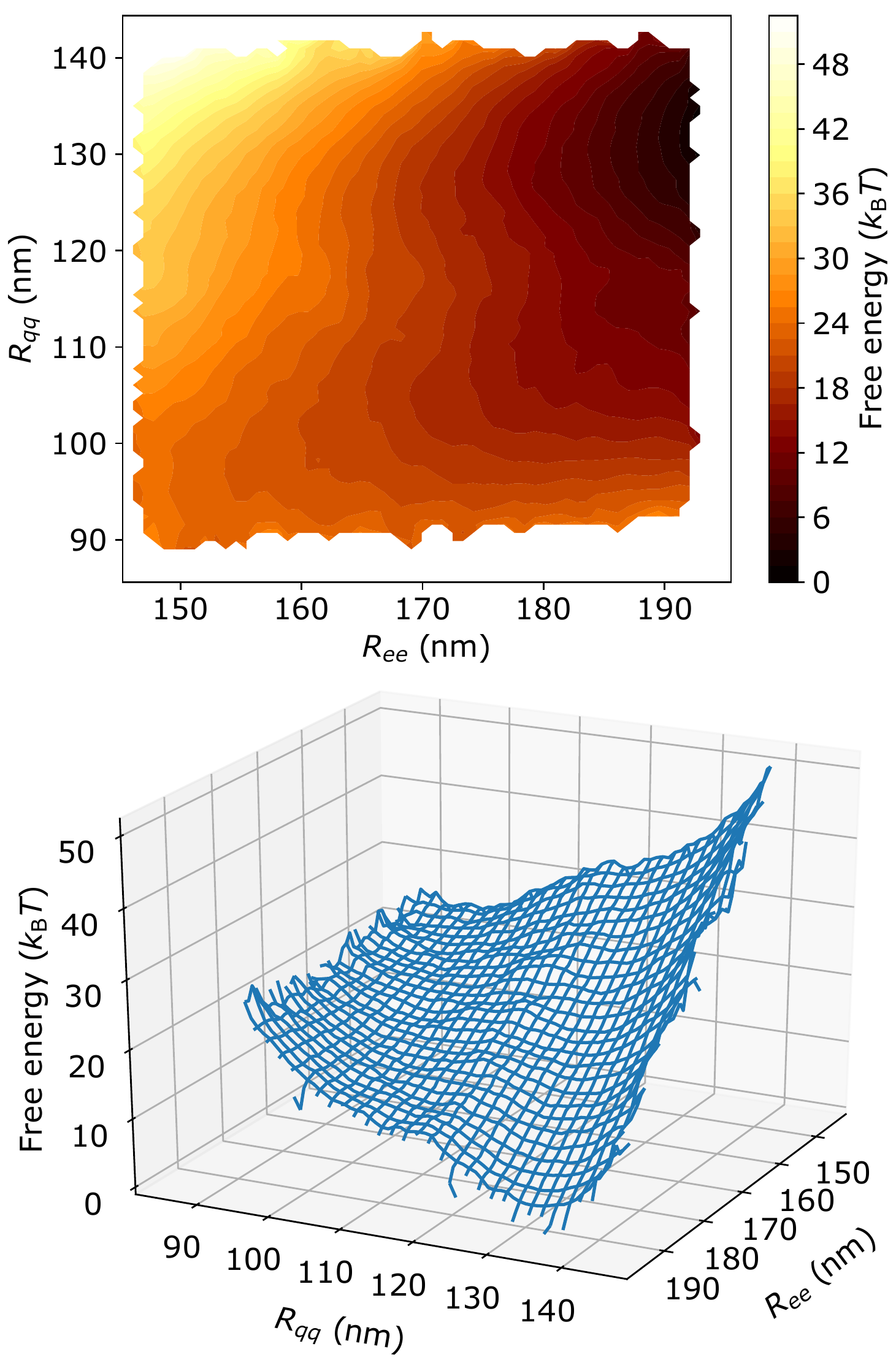}
    \caption{Two-dimensional free-energy landscape of design 1 as a function of $R_{ee}$ and $R_{qq}$
    The two plots provide two different representations of the same landscape.}
    \label{fig:A(r)_2d}
\end{figure}

The relative smoothness of the free-energy landscape in the two regimes is indicative that the landscape is well sampled in these regimes. However, accurately locating the point at which the crossover between the two regimes occurs is more challenging, because kink formation and resolution are relatively slow processes on the simulation time scales.
One way to aid the sampling in such cases is to perform two-dimensional umbrella sampling where a second order parameter is introduced that can differentiate between the two forms. The order parameter that we introduce is $R_{qq}$
the distance between points one quarter and three quarters
along the nanotube; for a given $R_{ee}$ this should have a smaller value in the kinked states because the nanotube sections either side of the kink are relatively straight.

The resulting two-dimensional free-energy landscape is shown in Fig.\ \ref{fig:A(r)_2d} for the transition region. It has two clear valleys corresponding to the kinked and homogeneously bent states, the latter having a steeper gradient. At intermediate values of $R_{ee}$ cuts through the landscape have two minima separated by a barrier as a function of $R_{qq}$.

Combining the information from the two-dimensional umbrella sampling in the transition region with the one-dimensional umbrella sampling at other values of $R_{ee}$ leads to the second free-energy profile shown in Fig.\ \ref{fig:A(r)}(b). Although the additional sampling has moved the transition to slightly larger $R_{ee}$, the relative change is small, suggesting that 1D sampling is sufficient to obtain a reasonably accurate representation of the free-energy landscape. 

\begin{figure}[t]
    \centering
    \includegraphics[width=70mm]{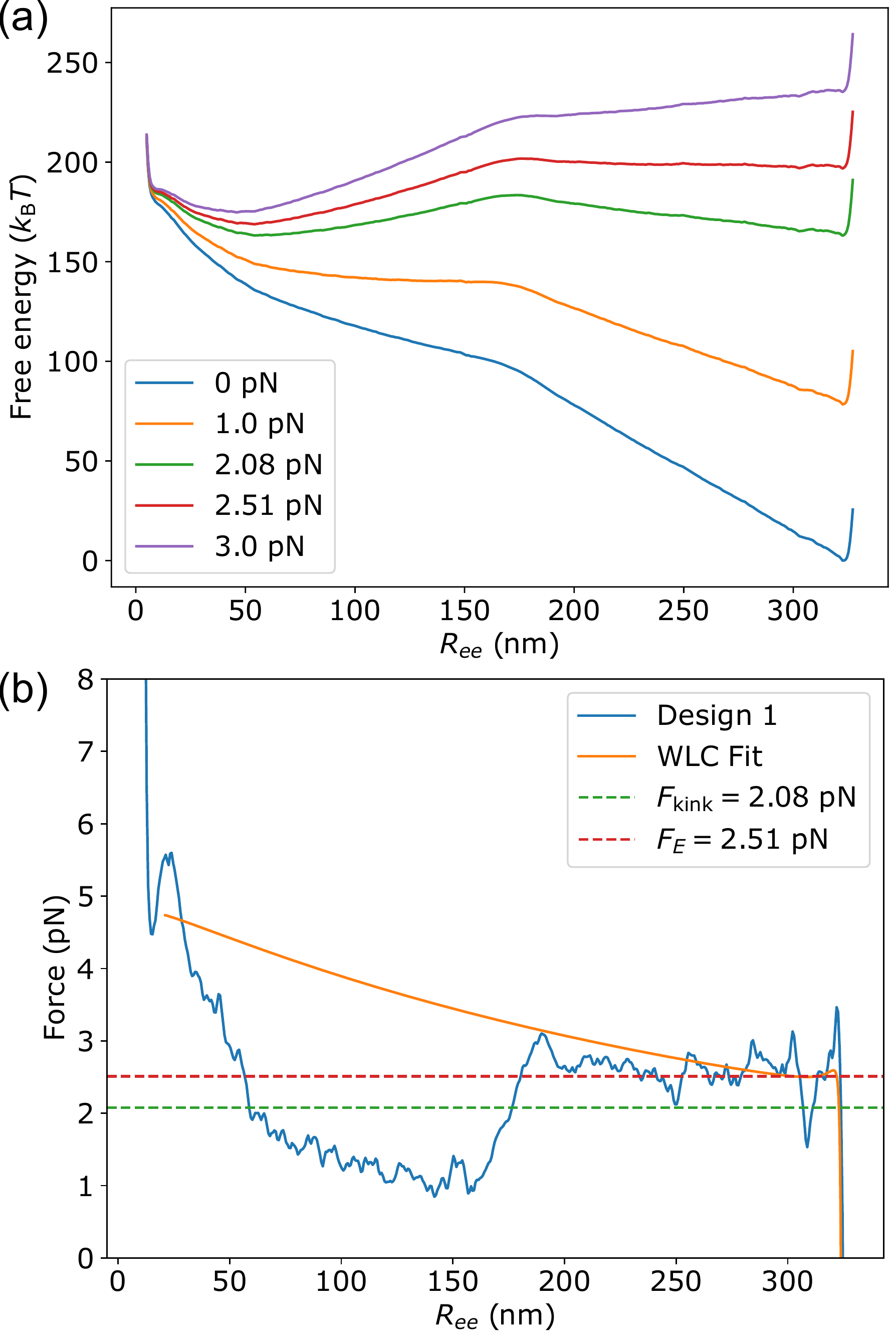}
    \caption{(a) The free-energy landscape for the design 1 nanotube at a series of compressive forces along the end-to-end vector, where $A(R_{ee};F)=A(R_{ee})+F R_{ee}$. 
    (b) $F(R_{ee})=-dA(R_{ee})/dR_{ee}$ for the computed landscape and its WLC fit.
    The two horizontal lines correspond to the force at which the two free-energy minima associated with the kinked and unkinked states are degenerate and the predicted Euler buckling critical force. Note that the noise in the force curve is due to a certain small degree of roughness in the simulation-derived landscapes.
    }
    \label{fig:force}
\end{figure}

We can also use our computed free-energy landscapes to help understand the response of the nanotube if a compressive force is applied along the end-to-end vector. For a macroscopic rod, Euler buckling predicts that the rod will remain straight until a critical force is reached, namely $F_E=k_\mathrm{B}T L_p \pi^2/L_c^2$
where $L_p$ and $L_c$ are the persistence and contour lengths, respectively. However, it is well known that in microscopic systems this instability is rounded by thermal fluctuations with the transition becoming less sharp as $L_c/L_p$ increases.\cite{Pilyugina17} For the current example $L_p/L_c\approx 20$, so the transition is expected to be quite sharp. 

In Fig.\ \ref{fig:force}(a) we show how the free-energy landscape depends on force 
and in Fig.\ \ref{fig:force}(b) we plot $F(R_{ee})=-dA(R_{ee})/dR_{ee}$ for both the computed landscape and the WLC fit. 
In line with the expectations outlined above, at $F\approx F_E=2.51$\,pN the section of the landscape associated with homogeneous bending becomes very flat and $R_{ee}$ decreases rapidly with increasing force. 

The other major feature is that the kinked state becomes a free-energy minimum as the force is increased. 
At $F=F_\mathrm{kink}=2.08$\,pN the free-energy minima associated with the kinked ($R_{ee}=54\,$nm) and unkinked ($R_{ee}=323\,$nm) states are degenerate separated by a barrier of $20.3\,k_\mathrm{B} T$; at equilibrium a transition to the kinked state would be expected at this force. At forces just above $F_E$ the kinked nanotube becomes the only stable state. 
Similarly, the plot of $F(R_{ee})$ has an unstable region of positive slope at $R_{ee}$ values associated with the crossover between the kinked and unkinked states, and at $F_\mathrm{kink}$ a Maxwell-like equal area rule applies in Fig.\ \ref{fig:force}(b).

The closest experimental realization of the above is the force-induced bending of DNA origami that has been achieved by single-stranded tethers that connect different parts of the origami. 
This approach has been used to bend six-helix bundles\cite{Liedl10} and cuboidal origami rods,\cite{Zhou14} where the degree of bending can be controlled by the length of the tether. 
However, in these examples, the forces exerted by the single-stranded tethers, rather than taking a constant value, also depend on the end-to-end distance.

\begin{figure*}[t]
    \centering
    \includegraphics[width=170mm]{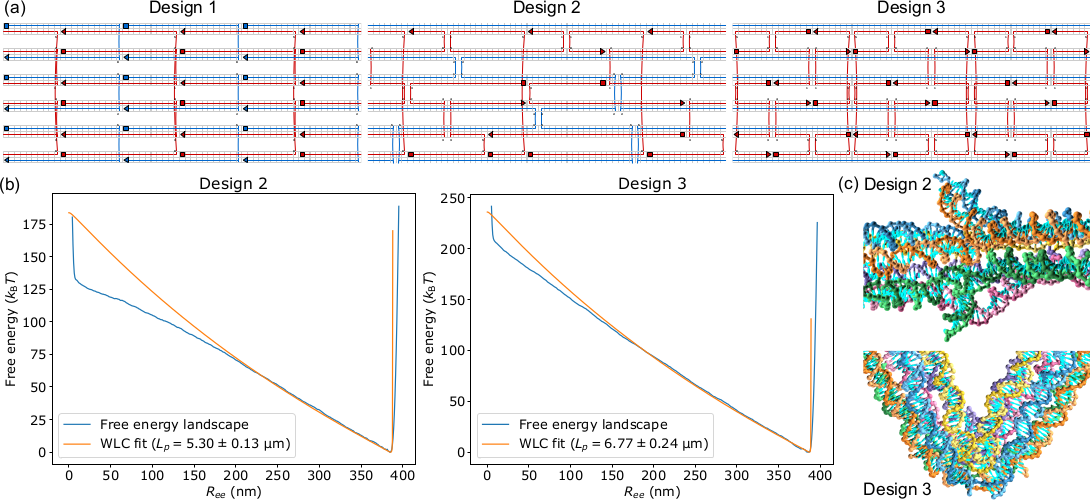}
    \caption{(a) Strand routing diagrams for sections of the three nanotube designs as represented in caDNAno.\cite{Douglas09} For designs 1 and 3 the pattern repeats along the nanotube. For the origamis (designs 2 and 3), the scaffold is in blue and staples in red. (b) Free-energy landscapes of designs 2 and 3 as a function of $R_{ee}$. (c) Close-up views (from the inside and side-on) of kinks for design 2 and 3 configurations with $R_{ee}=92$\,nm and 54\,nm, respectively. Further configurations of the two nanotubes are depicted in Figs.\ S5 and S6.}
    \label{fig:cadnano}
\end{figure*}

Comparing the free-energy landscapes of the three designs that we consider (Fig.\ \ref{fig:A(r)} and \ref{fig:cadnano}), although they all show the same basic behaviour, there are also significant variations between them. Design 1 is actually not an origami, but a single-stranded tile structure 
that is built from just six repeating 42-bp strands.\cite{Schiffels13} More important than the absence of a scaffold strand for the current study is that all the junctions are single crossovers (i.e.\ just one strand crosses between the adjacent helices) as this potentially allows significantly more freedom of motion at the junctions in response to stress. Design 2 is an experimentally-realized origami\cite{Pfitzner13,Siavashpouri17} that involves a majority of double crossovers. Notably, the scaffold crossovers are located very close to staple crossovers, with small 2-bp domains separating them; the dissociation of these 2-bp sections can lead to significantly increased local flexibility. Design 3 was constructed as a contrast to design 2 with the aim of increasing the mechanical strength. In this origami all crossovers are mediated by staples and all are double crossovers. All nicks are placed midway between junctions meaning the shortest domain is 7 bp; this also leads to planes where every helix has either a junction or a nick, which are likely to be preferred sites for local bending.

The persistence lengths obtained from the WLC fits for designs 1 and 2 are 
$\SI{6.5}{\um}$ and $\SI{5.3}{\um}$; these agree with the values extracted from the thermal fluctuations of the nanotubes in unbiased simulations.\cite{Chhabra20}
However, they are higher than the experimental values of \SI{3.3}{\um} \cite{Schiffels13} and \SI{2.4}{\um} \cite{Siavashpouri17}. We note that for design 1, there was a significant dispersion of persistence lengths (\SI{1.6}{\um} to \SI{4.7}{\um}) between the individual nanotubes probed in the experiments, which was attributed to structural differences among different nanotubes \cite{Schiffels13}. 
Experimental persistence length have also been found to depend on the post-assembly treatment of the origamis.\cite{Liedl10}
This suggests that the experiments are measuring the persistence lengths of a heterogeneous ensemble of nanotubes that include some with structural defects that lower the persistence length. By contrast, the simulations measure the persistence length of a perfectly assembled nanotube. That oxDNA overestimates the stretch modulus may also contribute to this difference.\cite{Chhabra20}

The persistence lengths of designs 1 and 3 are very similar, but for design 2 it is lower by about 20\%. If one inspects the number of intact base pairs as a function of $R_{ee}$ (Fig.\ S7), there is a noticeable difference between the designs; while for designs 1 and 3 there are no significant changes until the onset of kinking, for design 2 there is a small but noticeable decrease. Inspection of snapshots of design 2 reveals that the short 2-bp domains noted above are occasionally broken; this is probably one reason for this nanotube's greater flexibility--- the less regular distribution of crossovers may also play a role.

Design 1 shows a clear change of slope of the free-energy landscape associated with kinking. However, the deviation from the worm-like chain fit is more gradual for the two origamis and starts to occur prior to clear kink formation (see the example configurations in Figs.\ S5 and S6 and the energetic and structural descriptors in Figs.\ S7 and S8). For example, for design 3, $R_{ee}$ has to be significantly less than 100\,nm for a kink to appear, even though deviations from the worm-like chain fit start to appear at about 150\,nm. This non-linear elasticity is because the nanotubes exploit the heterogeneity of their local mechanical properties to partially localize stronger bending at mechanically more flexible regions.

The kinks in the nanotubes look somewhat like the kinks in a macroscopic tube. The cross-sections become flattened and wider, but the detailed structure depends on the nanotube design, the main difference being how the tubes respond to the compressive stresses on the inside of the kink. The single crossovers in design 1 mean the helical sections have the most freedom to reorient, just through loss of the coaxial stacking interactions at the junction and perhaps some fraying of adjacent base pairs. In the example in Fig.\ \ref{fig:A(r)}(d) this allows the outer two helices in the inside of the kink to splay out, and in the (orange) middle helix a section is oriented almost perpendicular to the nanotube axis. By contrast the three helices on the outside of the kink are almost co-planar and bend continuously around the kink. 

In design 2, this freedom of motion is achieved by the loss of base pairs, particularly associated with the short 2-bp domains already mentioned, but also by the unbinding of short staple end-domains. In design 3, the helical sections are most constrained, because of the exclusively double crossovers and the lack of short staple domains. Kinking occurs at the planes of nicks and junctions by the loss of coaxial stacking, but the free-energetic advantage of kinking rather than homogeneously bending is least for this system. The design has significantly increased the nanotube's stiffness and strength with respect to extreme bending compared to design 2. 

The increased resistance to kinking also leads to differences in the predicted behaviour when a compressive force is applied along the end-to-end vector compared to design 1 (Fig.\ S9). For design 2, the kinked state becomes equal in free energy to the unkinked state only very close to the Euler buckling critical force, whereas for design 3 the force-dependent free-energy landscape only ever has a single free-energy minimum and
the nanotube is instead predicted to transition continuously between unkinked and kinked states.

\begin{figure}[t]
    \centering
    \includegraphics[width=84mm]{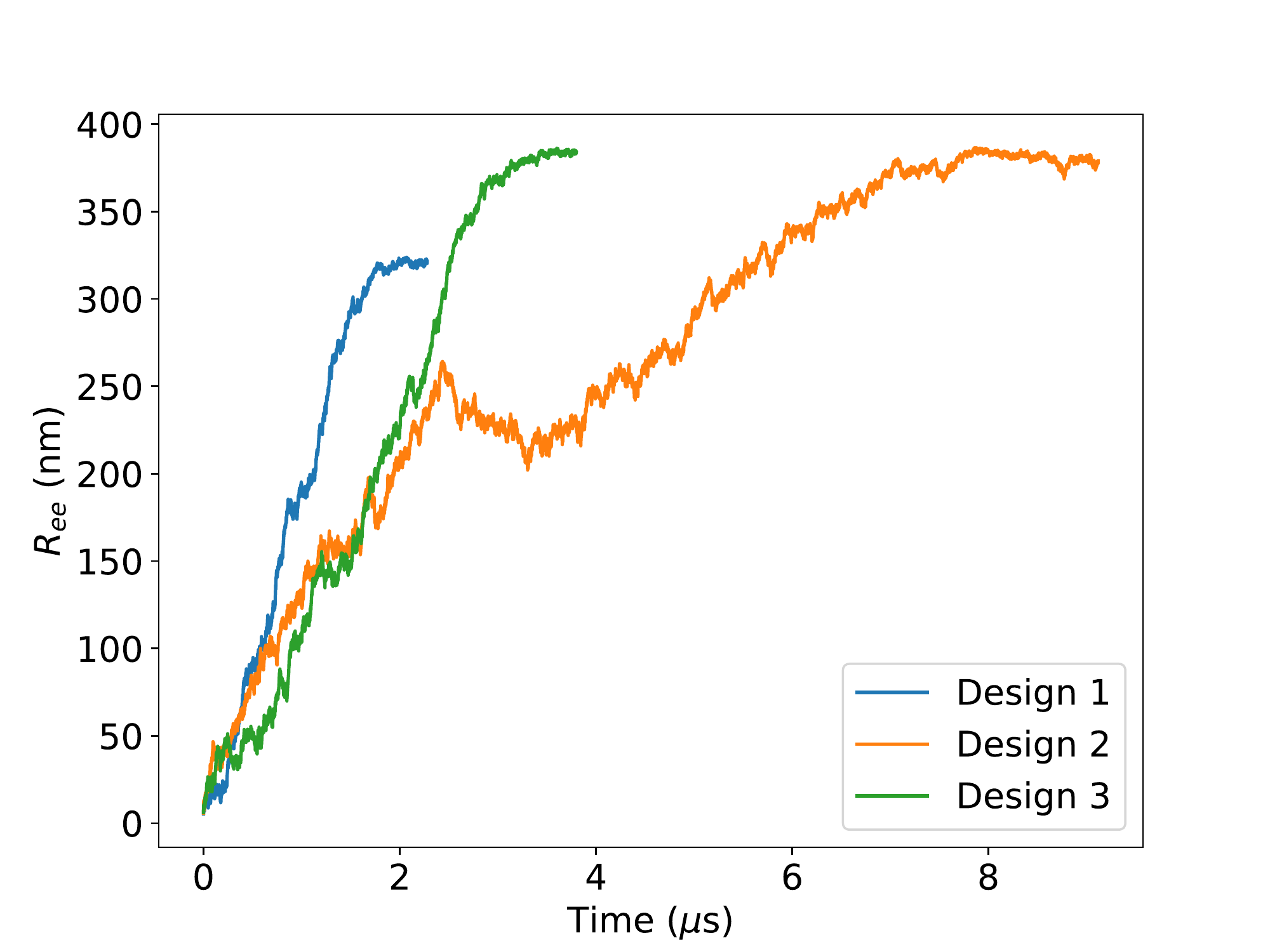}
    \caption{Time evolution of $R_{ee}$ in unbiased molecular dynamics simulations that start from nanotube configurations with $R_{ee}$ close to zero that were generated in our umbrella sampling simulations. All are able to relax back to their equilibrium state. Movies of these trajectories are included as Supporting Information.}
    \label{fig:r_release}
\end{figure}

It is also of interest to test whether extreme bending leads to any irreversible damage to the nanotubes. One reason for irreversibility would be if the stresses lead to staple dissociation, but such events were never seen in any of our simulations. To explore this further, we took configurations from the most bent umbrella sampling windows, and used them as starting points in unbiased molecular dynamics simulations. The time evolution of the end-to-end distance is shown in Fig.\ \ref{fig:r_release} for all three designs. All were able to relax back to their equilibrium $R_{ee}$ value. Furthermore, after they have fully relaxed, the number of pairs of interactions is essentially the same as that of a normal unstressed nanotube (Table S1). This suggests that strong bending and kink formation do not induce irreversible structural changes to the nanotubes.

The effects of the design differences are also evident from these simulations. In particular, for design 2 the slow reannealing of broken base pairs, especially at the site of the kink, led to a slower recovery of the equilibrium structure. 

\subsubsection{DNA nanotube stretching}

\begin{figure}[t]
    \centering
    \includegraphics[width=84mm]{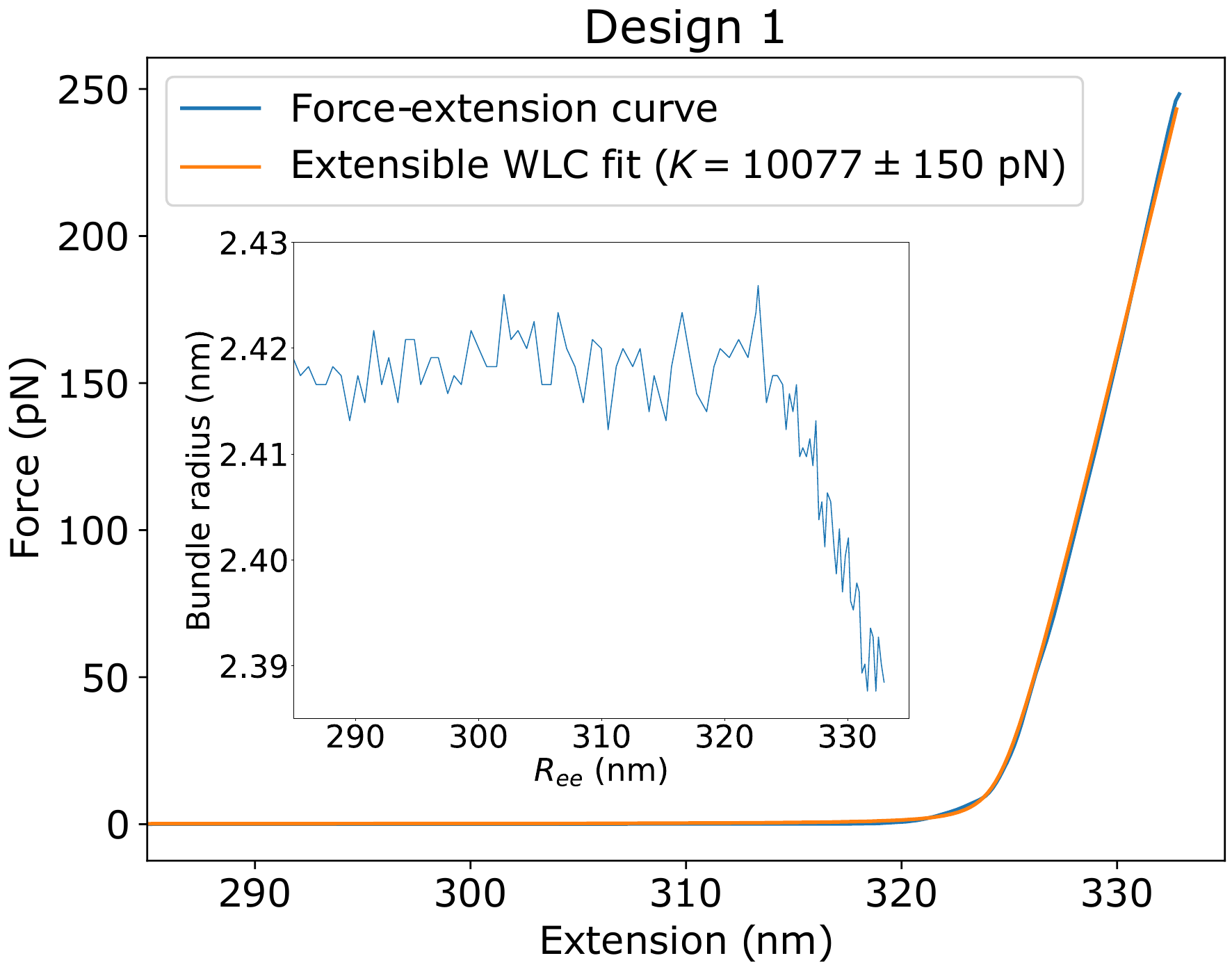}
    \caption{Force-extension curve for the design 1 nanotube along with the extensible WLC fit (see Table S3 for the fit parameters). The inset shows the average nanotube radius as a function of $R_{ee}$.}
    \label{fig:F(z)}
\end{figure}

The form of the free-energy landscape at $R_{ee}$ values beyond the minimum contains information about the stretching of the nanotubes. In particular, the force-extension curve of the nanotube can be extracted from $A(R_{ee})$ (as detailed in the Supporting Information); then by fitting this curve to that for an extensible WLC the stretch modulus can be obtained.\cite{Odijk95} 
The resulting force-extension curve and the fit to it are shown in Fig.\ \ref{fig:F(z)} for design 1 and in Fig.\ S10 for the other two designs. All designs give a value of the extensional modulus that is roughly 10\,000\,pN. This is significantly less than the six times the oxDNA extensional modulus for double-stranded DNA (2700\,pN) that one might have expected. The reason is that the nanotube can be extended not only by the stretching of the individual helices, but also by a straightening of the helices within the origami structure. It is well evidenced that the helices in a relaxed origami are not perfectly parallel, but splay out by a small angle at each junction.\cite{Rothemund06,Baker18,Snodin19} Reduction of this angle will cause a reduction in the interhelix distance and hence the nanotube radius. Such a reduction in the nanotube radius is clearly seen in the stretching regime, whereas it is roughly constant in the regime associated with homogeneous bending (Fig.\ \ref{fig:F(z)} inset). 

\subsection{Twisted DNA origami sheets}

As mentioned in the introduction, when an inextensible sheet is twisted there is an inevitable coupling to bending, as is illustrated for an A4 sheet of paper in Supplementary Fig.\ S11. However, the curvature that results can be of either sign. In the limit that this curvature is localized at a line (i.e.\ a fold), the two states correspond to a ``valley'' fold about one diagonal or a ``mountain'' fold about the opposite diagonal. 

Does a similar picture hold for a twisted 2D origami sheet? Although such origamis are not inextensible, the bending modulus perpendicular to the directions of the helices is expected to be relatively low. For example, it was not possible to get a well-defined cryo-EM structure for a 2D origami sheet because the fluctuations were so large.\cite{Kube20}. Furthermore, SAXS experiments suggest that twisted sheets exhibit significant curvature, with CanDo\cite{Kim12,Castro11} modelling predicting that the curvature increases with twist.\cite{Baker18} 

Because there is no difference between the two sides of a paper sheet, the two states for the twisted sheet are equivalent. By contrast, the two faces of a chiral DNA origami sheet are inequivalent. However, it is not yet known whether both states are possible for an origami sheet, or whether instead one is overwhelmingly favoured? For example, the SAXS measurements are not sensitive to the sign of the curvature. AFM studies have attempted to infer the preferred curvature based on the assumption that the origami are more likely to bind to the surface when they land convex face down.\cite{Marchi14,Stammers18}

We first explored these questions for a 2D origami sheet with a Rothemund-like design; such single-layer origamis are right-twisted because the junction spacing in the design would require a DNA pitch length of 10.67 base pairs to be flat; this is larger than the natural pitch length of about 10.5 base pairs. 
On relaxing the initial flat configuration generated from converting the caDNAno design, the origami twisted and rolled up as expected (this effect has been seen before for oxDNA\cite{Khara18}). However, when we repeated this relaxation multiple times, the resulting structures exhibited both curvatures. These results indicated that both forms are free-energy minima and that the initial flat geometry is in the transition region with stochastic effects during the relaxation causing the system to fall into one or other of the two basins. 

\begin{figure*}[ht!]
    \centering
    \includegraphics[width=170mm]{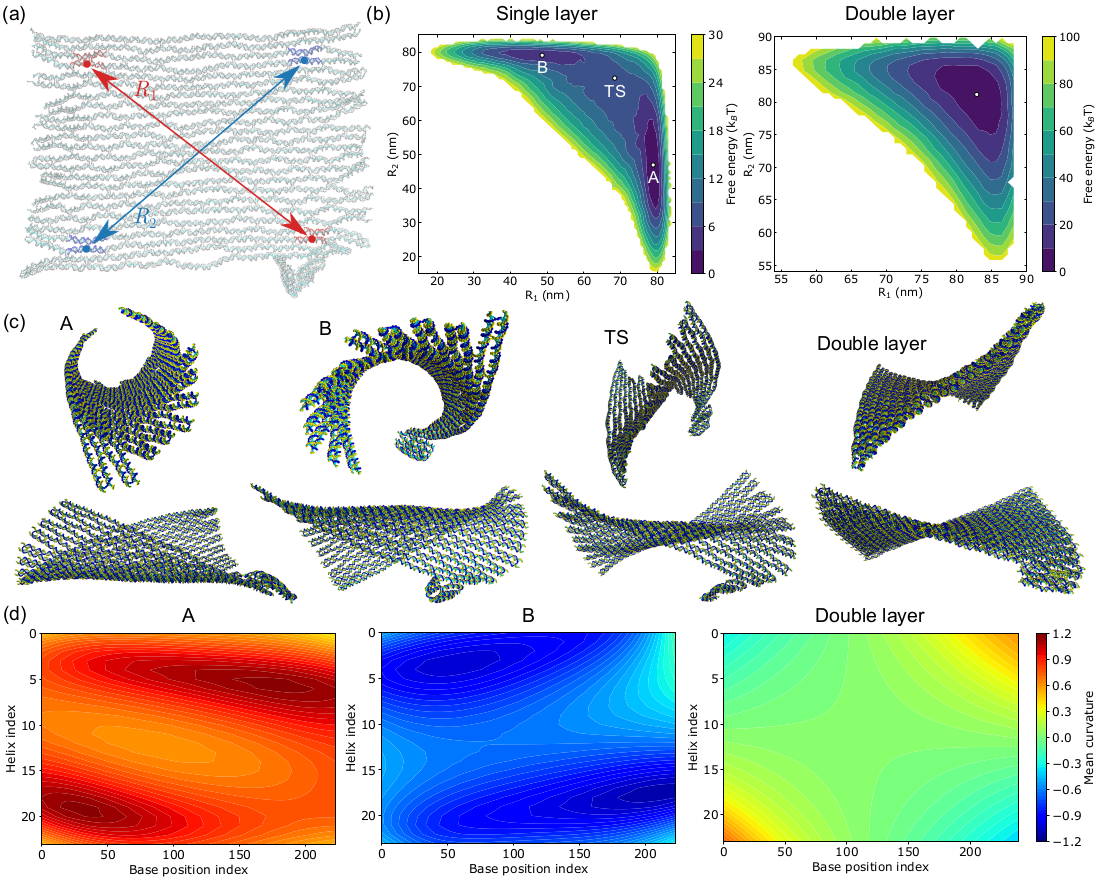}
    \caption{(a) The two diagonal distances $R_1$ and $R_2$ that are used as order parameters in the umbrella sampling illustrated for a flat configuration of the single-layer sheet. 
    The additional short helices in the bottom right corner are to aid identification of the orientation of the origami. Positive curvature corresponds to the edges of the sheet being curved up out of the plane.
    (b) Free-energy landscapes of the single-layer and double-layer twisted sheets. (c) End-on and side-on views of the configurations associated with the two free-energy minima and the transition state of the single-layer sheet, and with the free-energy minimum of the double-layer sheet. (d) Mean curvature (the average of the principal curvatures) of configurations at the two free-energy minima of the single-layer sheet and the free-energy minimum of the double-layer sheet as a function of the base-pair position within the sheet. 
    }
    \label{fig:sheet}
\end{figure*}

To characterize this bistability further we computed the free-energy landscape of the 2D origami sheet as a function of two order parameters, namely the two diagonal distances $R_1$ and $R_2$ that are shown in Fig.\ \ref{fig:sheet}(a). This was again achieved using umbrella sampling with each individual simulation centred around a different grid point in the 2D order parameter space. 

The resulting free-energy landscape is shown in Fig.\ \ref{fig:sheet}(b). The landscape has two free-energy minima and is roughly symmetric about the diagonal $R_1=R_2$. For each free-energy minimum one diagonal distance is roughly unperturbed (with a value of about 80\,nm) whilst the other diagonal distance is reduced by about 40\% due to the curvature. The free-energy minima are also very asymmetric with the soft mode corresponding to changing the degree of curvature perpendicular to the ``unperturbed diagonal''. 

Inspection of the configurations associated with the two free-energy minima confirms that both have a right-handed twist but that they have opposite curvatures. Quantitative confirmation comes from the plots of the mean curvature in Fig.\ \ref{fig:sheet}(d). The mean curvature for the two forms is always positive or negative, respectively, with the pattern of local variation showing an approximate symmetry with respect to the relevant diagonal.

There is a small free-energy difference between the two minima of $3.5\,k_\mathrm{B}T$. Although there is no reason to expect the two free-energy minima to be degenerate because they are not equivalent by symmetry, the physical reason why the positive curvature form is slightly more stable is not clear. The free-energy barrier (measured with respect to the more stable form) is $7.5\,k_\mathrm{B}T$. Although this is sufficiently high that transitions would be unlikely to be seen in unbiased simulations, on typical experimental time scales both forms would likely be sampled. 

The free-energy transition state occurs near to the $R_1=R_2$ diagonal at $(R_1,R_2)=(68\,\mathrm{nm},72\,\mathrm{nm})$, but slightly displaced towards the higher free-energy minimum. It is not a flat sheet but corresponds to a more saddle-like configuration that retains the twist but has an overall mean curvature that is close to zero. 

The second example that we have considered is a twisted two-layer origami based upon a square lattice of helices. As observed experimentally,\cite{Zhang20} this second layer significantly increases the stiffness of the origami compared to the single-layer designs; in particular we expect the sheet to be much stiffer with respect to bending perpendicular to the helix axis, as bending is now coupled to the stretching and compression of the inter-helix bonds on the outside and inside of the bend, respectively. 

The free-energy landscape of the twisted two-layer origami is quite different from the single-layer origami. It has a single free-energy minimum with $R_1\approx R_2$. The increased bending stiffness has removed the bistability that is observed for the single-layer origami. A saddle-like geometry, which was characteristic of the transition region in Fig.\ \ref{fig:sheet}(b), is now lowest in free energy, as also predicted by CanDo.\cite{Zhang20}. In this geometry much of the stress due to twist is accommodated through stretching and compression of the origami rather than just by bending. Reflecting the origami's greater stiffness, the free energy increases much more rapidly on moving away from the minimum. Hence, the magnitude of the structural fluctuations will be much less than for a single-layer origami sheet. 

The mean curvature for a perfect saddle would be zero due to the cancellation of the negative and positive curvatures along the two diagonals. Although not exactly zero, the mean curvature for the two layer origami is close to zero over the whole origami except for very close to the corners. The Gaussian curvature (the product of the principle curvatures), by contrast would be expected to be uniformly negative, as it is, except again at the corners (Fig.\ S12).

\section{Conclusions}

We have shown that it is feasible to compute the free-energy landscapes associated with the mechanical deformation of large DNA nanostructures, such as DNA origami, when described with a nucleotide-level coarse-grained model. The use of umbrella sampling with many windows allows all parts of the free-energy landscapes to be sampled in a highly parallel manner.

The first example that we considered is the extreme bending of DNA nanotubes, for which we compute the free-energy landscape as a function of an end-to-end distance. Such nanotubes are extremely stiff and are promising building blocks for nanoscale engineering. In the homogeneous bending regime, the behaviour is well described by the worm-like chain model. In the extreme bending regime the bending stress localizes at one or two kinks, where the ease of kinking and the nature of the deformations at the kink depends sensitively on the local features of the nanotube design. Importantly, the nanotubes can also reversibly recover from this kinking when the stress is removed. The landscapes at end-to-end distances larger than the nanotube's relaxed length also allows us to extract the stretch modulus of the nanotubes. Their values are somewhat less than would be expected for six helices stretched in parallel, because the nanotubes can also respond to the tension by reducing their radius.

We have also characterized the free-energy landscapes of twisted DNA origami sheets. For a single-layer sheet there are two nearly degenerate free-energy minima because of the strong coupling of the twist to the soft bending mode perpendicular to the helices with the two minima having opposite curvature. This bistability disappears for a twisted two-layer DNA origami; the stiffening of the bending mode means that the preferred form has a saddle-like geometry where the twist stress is also accommodated through stretching and compression of the sheet. 

In order for it to be feasible to compute free-energy landscapes for DNA origamis using the current approach it is important that the properties of the systems vary relatively continuously as the order parameters change, so that equilibration within each umbrella sampling window can be relatively rapid. Such calculations becomes more problematic when discontinuous changes occur. We saw this for the kinking transition in the nanotubes, where to more accurately locate the transition we needed to introduce a second order parameter that provided a more continuous pathway between the kinked and unkinked states. Thus, the approach can be most easily applied to landscapes associated with mechanical deformations. By contrast, calculating free-energy landscapes that involve significant assembly or disassembly, although feasible for small and medium-sized\cite{Harrison19,Mosayebi15,Romano13,Ouldridge13,Ouldridge10,Khara18} DNA systems, is likely to be computationally very challenging for systems of the size of DNA origamis.

As well as being used to explore the fundamental mechanical properties of DNA origami, we expect the current methods to find application in DNA mechanotechnology\cite{Blanchard19} where DNA origami systems are increasingly being used as part of force-sensing\cite{Dutta18} or force-applying\cite{Nickels16b} devices, and an accurate characterization of the compliance of the origami devices under the relevant mechanical loads is likely to be important for their calibration. Accurate characterization of origami mechanics is also likely to be important for understanding stress accumulation in the self-assembly of origamis into higher-order structures, be it an unwanted side effect which hinders correct assembly, or a design strategy to control and direct the assembly.\cite{Berengut20,Hagan20}

\section{Methods}

To model the DNA nanostructures we use the oxDNA coarse-grained model.\cite{Ouldridge11,Sulc12,Snodin15}  In the oxDNA model, each nucleotide is represented by a rigid body. The nucleotides interact through a pairwise potential which has terms representing backbone connectivity, excluded-volume interactions, hydrogen bonding between base pairs, stacking interactions, and electrostatic interactions between backbone phosphate groups. Solvent is modelled implicitly as a dielectric continuum, and we choose to use a salt concentration of [Na$^+$]\,=\,1.0\,M, which is representative of the high salt conditions typically used for DNA nanotechnology.

The oxDNA simulation code was used to perform molecular dynamics simulations in the canonical ensemble, in particular making use of its GPU implementation.\cite{Rovigatti15} In the umbrella sampling simulations, harmonic biasing potentials in the order parameters were placed at the centre of each window. The free-energy landscapes were constructed from the order-parameter probability distributions for each window using the weighted-histogram analysis method (WHAM).\cite{Kumar92} For the 1D umbrella sampling of the DNA nanotubes 500 windows were used, and in the 2D umbrella sampling of the origami sheets 10\,000 windows were used.
We iterated the sets of umbrella sampling simulations until the WHAM-generated landscapes no longer changed.

Additional simulation and analysis details can be found in the Supporting Information.

%\begin{suppinfo}
%%{\bf Supporting Information} 
%The Supporting Information is available free of charge on the ACS Publications 
%website at DOI: .
%\begin{itemize}
%\item Further details of simulation methods and additional supporting results (PDF)
%\item Unbending animations (MPG)
%\end{itemize}
%\end{suppinfo}

\bigskip
{\bf Conflict of Interest:} The authors declare no competing financial interest.

\begin{acknowledgement}
CKW is grateful to the Croucher Foundation for financial support.
The authors acknowledge the use of the University of Oxford Advanced Research Computing (ARC) facility (http://dx.doi.org/10.5281/zenodo.22558) and the resources provided by the Cambridge Service for Data Driven Discovery (CSD3). 
We are grateful to the group of Christoph W\"{a}lti for helpful discussions.
\end{acknowledgement}

%\bibliography{biblio}

\providecommand{\latin}[1]{#1}
\makeatletter
\providecommand{\doi}
  {\begingroup\let\do\@makeother\dospecials
  \catcode`\{=1 \catcode`\}=2 \doi@aux}
\providecommand{\doi@aux}[1]{\endgroup\texttt{#1}}
\makeatother
\providecommand*\mcitethebibliography{\thebibliography}
\csname @ifundefined\endcsname{endmcitethebibliography}
  {\let\endmcitethebibliography\endthebibliography}{}

\clearpage
\setcounter{figure}{0}
 \makeatletter
 \renewcommand{\thefigure}{S\@arabic\c@figure}
 \setcounter{equation}{0}
 \renewcommand{\theequation}{S\@arabic\c@equation}
 \setcounter{table}{0}
 \renewcommand{\thetable}{S\@arabic\c@table}
 \setcounter{section}{0}
 \renewcommand{\thesection}{S\@arabic\c@section}
\section{Supporting Information}

\subsection{Additional methodological details}

\begin{figure}[t]
    \centering
    \includegraphics[width=84mm]{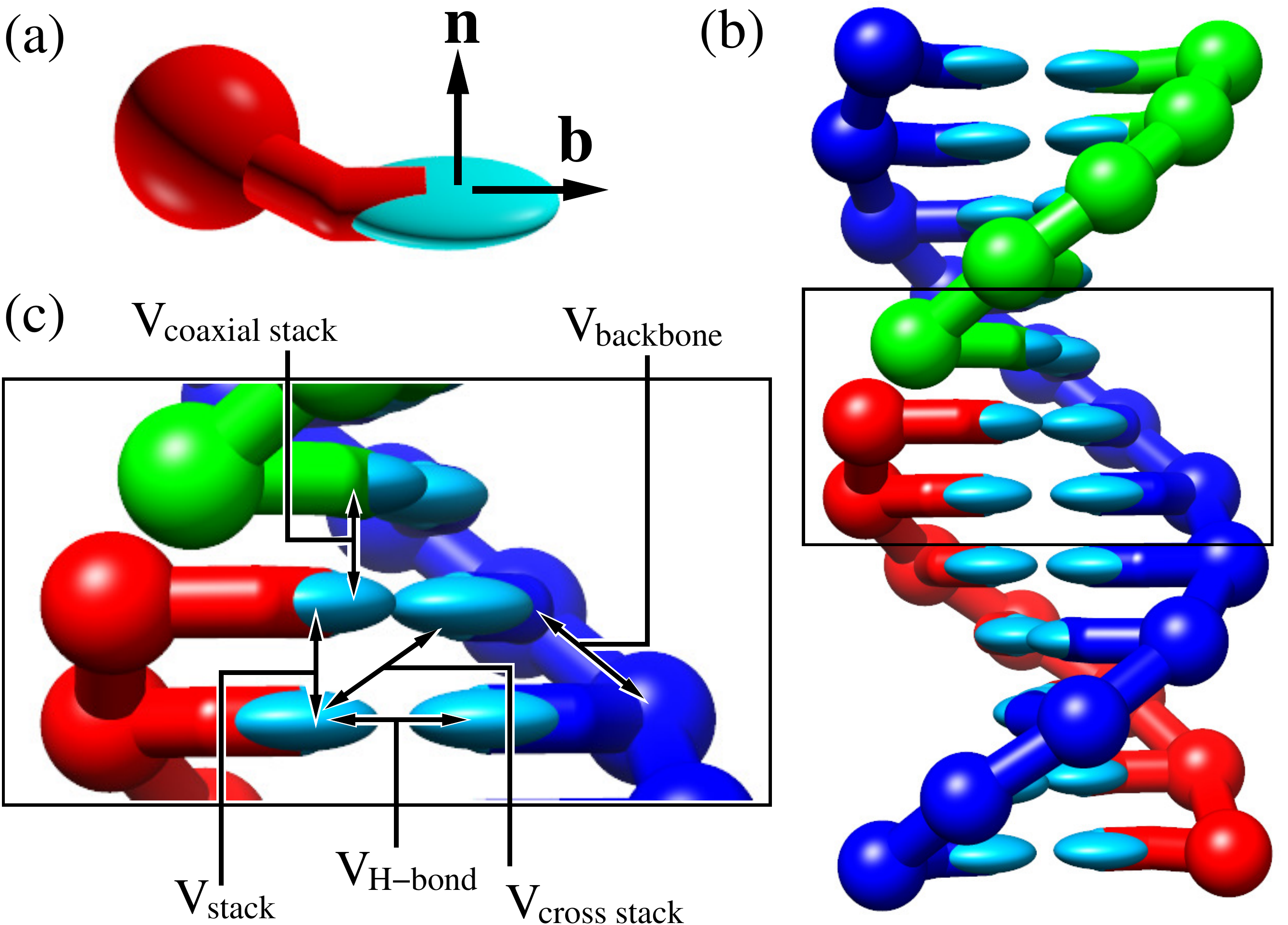}
    \caption{(a) An oxDNA nucleotide along with the ``base'' and ``normal'' vectors used to define its orientation. (b) A nicked double helix with 12 base pairs. (c) A close-up of the double helix illustrating some of the interactions in the oxDNA model.}
    \label{sfig:oxDNA2}
\end{figure}

\paragraph{oxDNA simulations}
We used the oxDNA coarse-grained model\cite{Ouldridge11,Sulc12} (Fig.\ \ref{sfig:oxDNA2}) to perform molecular dynamics simulations of the DNA nanostructures. We used the second-generation version of the model (sometimes called ``oxDNA2'');\cite{Snodin15} one of the improvements in this version of the model was a tuning of the potential parameters to better reproduce the structures of large DNA nanostructures. As we are concerned here with the fundamental properties of the free-energy landscapes and not any potential sequence dependence to them, we used the sequence-averaged version of the model in which the strength of the interactions are independent of the identity of the nucleotides involved (note base pairing still can only occur between complementary nucleotides).

Simulations were performed at 300\,K using a Langevin thermostat.
The time-step used was 0.005 in the internal simulation units of the oxDNA code, which corresponds to 15\,fs. The solvent environment is treated implicitly as a dielectric continuum. We use a salt concentration of [Na$^+$] = 1.0\,M, which provides a reasonable representation of the high salt conditions typically used for DNA nanotechnology. Due to large system sizes, we use the GPU implementation of the oxDNA simulation code.\cite{Rovigatti15}

To apply umbrella sampling in molecular dynamics simulations it is necessary to compute the contributions to the forces arising from the umbrella potential. This is straightforward for the continuous distance-based order parameters and the harmonic umbrella potentials that we use here.

\paragraph{Generating starting configurations}
We converted the caDNAno design files of the origamis into oxDNA format using the tacoxDNA package \cite{Suma19}. The converted configurations cannot serve as starting configurations for molecular dynamics simulations because of nucleotides experiencing large forces due to particle overlaps or extended bonds. Therefore, the potential energy of these configurations is first minimized for 200 steps using a steepest-descent algorithm, and then the configurations are relaxed in a molecular dynamics simulation using a modified backbone potential for $10^6$ steps. After that, the extended bonds have typically returned to their normal lengths, and the configurations are ready for simulation using the standard oxDNA force field. All designs were equilibrated for a further $10^8$ steps, corresponding to about \SI{1.5}{\us}.

\subsection{Nanotube}
\begin{figure*}
    \centering
    \includegraphics[width=\textwidth]{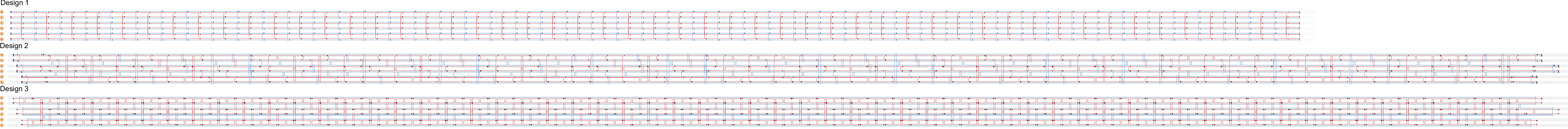}
    \caption{Full caDNAno designs for the three DNA nanotubes.}
    \label{sfig:cadnanoSI}
\end{figure*}

\paragraph{1D Umbrella Sampling}
We used umbrella sampling \cite{Torrie77} to calculate the free-energy landscape as a function of the order parameter $R_{ee}$, the end-to-end distance of the nanotube. To avoid ``end'' effects associated with the greater splaying of the helices at the nanotube ends and their larger fluctuations, we defined the end-to-end distance as the distance between the centres of mass of two groups of nucleotides that are slightly in from the ends. For design 1, the two groups contain the nucleotides at positions 44--64 and 1010--1030 respectively. For designs 2 and 3, the two groups contain the nucleotides at positions 49--69 and 1204--1224 respectively. These nucleotides are highlighted in Fig.\ 1(a). The nucleotide positions are as defined in the caDNAno design (Fig.\ \ref{sfig:cadnanoSI}).

For each nanotube design, we performed a set of simulations where the order parameter was restrained with a harmonic potential in each sampling window. The range was from $R_{ee} = 0$ to $R_{ee} \approx 1.02\, L_{c}^\mathrm{approx}$, where $L_{c}^\mathrm{approx}$ is an estimate of the contour length of the nanotube that is obtained by multiplying the number of base pairs along the nanotube between the two centres of mass of the groups of nucleotides by the rise per base pair (\SI{0.34}{nm}). 

\begin{figure*}[t]
    \centering
    \includegraphics[width=170mm]{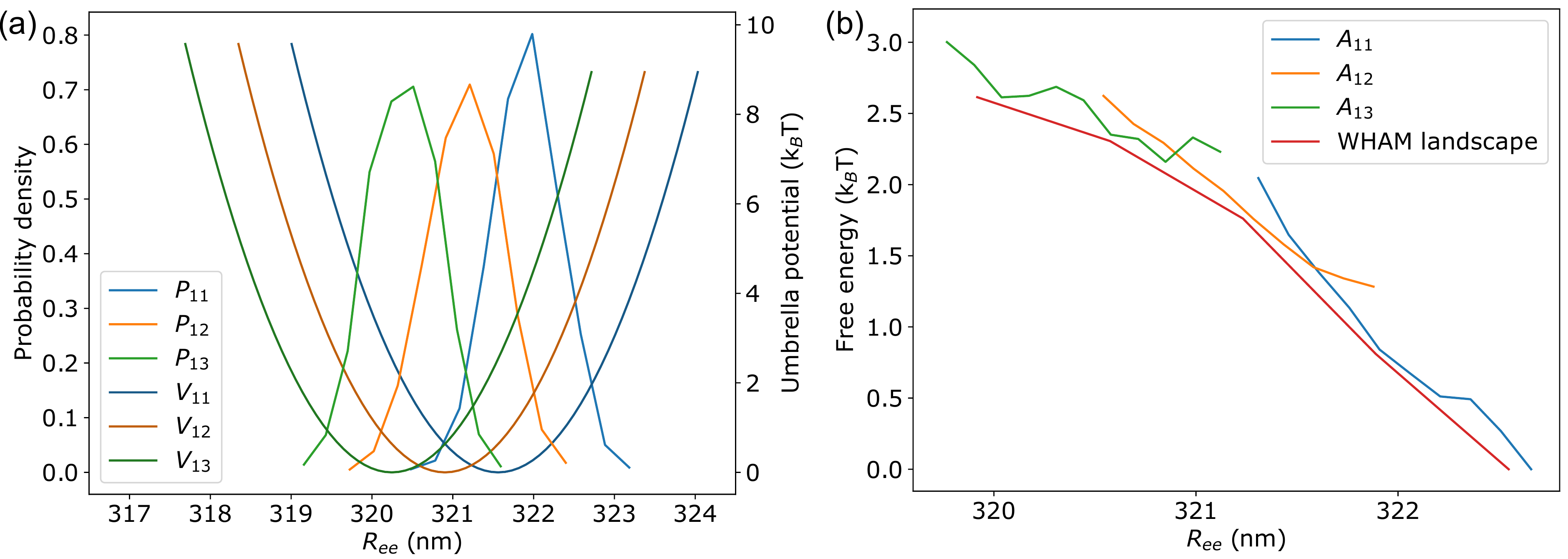}
    \caption{(a) Examples of the umbrella potentials $V_i(R_{ee})$ used for three windows for the bending of a DNA nanotube (design 1), and the probability distributions for those windows $p_i(R_{ee})$. $i$ is the index of the umbrella sampling window. (b) From each probability distribution an estimate of the free-energy $A_i(R_{ee})$ can be obtained that will be most accurate for those $R_{ee}$ values that are most well sampled in that simulation. If there is significant overlap between adjacent probability distributions an accurate best estimate of the overall free-energy landscape can be obtained using WHAM, as illustrated.}
    \label{sfig:US_example}
\end{figure*}

Before performing the umbrella sampling simulations, we first prepared starting configurations for each window using non-equilibrium ``pulling'' simulations. Starting from an equilibrated configuration restrained at $R_{ee} = L_{c}^\mathrm{approx}$ with a harmonic bias potential of stiffness $k=\SI{57.09}{pN/nm}$, the bent configurations were generated by gradually reducing the equilibrium position of the harmonic bias potential at a constant rate such that it reached zero after $10^8$ steps. Configurations were outputted every $2\times10^5$ steps, resulting in a total of 500 configurations that were then used as the starting configurations for the same number of equally-spaced windows from $R_{ee} = 0$ to $R_{ee} = L_c^\mathrm{approx}$. Similarly, for the stretched configurations, the equilibrium position of the harmonic potential was gradually increased at the same constant rate as above, until the nanotube was no longer stable. Configurations were also outputted every $2\times10^5$ steps, resulting in a total of 28, 27, and 37 configurations for designs 1, 2 and 3, respectively, that were used as the starting configuration of the same number of equally-spaced umbrella sampling windows.

The above configurations were then used as starting points for simulations in which the configurations were constrained by a harmonic umbrella potential of stiffness $k_{ee}=\SI{17.12}{pN/nm}$ centred at the value of $R_{ee}$ corresponding to that window. The value of $k_{ee}$ was chosen so that the probability distributions of $R_{ee}$ for adjacent windows had significant overlap (Fig.\ \ref{sfig:US_example}). Each window was equilibrated for $10^6$ steps before a production run of $10^7$ steps. $R_{ee}$ was outputted every $10^3$ steps, giving $10^4$ data points for each window. Using the biased probability distributions of $R_{ee}$ in each window, we used the Weighted Histogram Analysis Method (WHAM) \cite{Kumar92,wham} to calculate the unbiased free-energy landscape of the system as a function of $R_{ee}$. 

The production runs were then repeated with the last configuration of the previous production run, and WHAM was performed on the new data to calculate the free-energy landscape again. This process was repeated until the new free-energy landscape was not statistically different from the previously calculated free-energy landscape. For the nanotubes considered here, 
four sets of production runs were needed for convergence.

\paragraph{2D Umbrella Sampling}
To improve sampling in the transition region between the kinked and unkinked states of the nanotube, two-dimensional umbrella sampling was performed for design 1. The order parameters were chosen to be $R_{ee}$, the end-to-end distance of the nanotube defined previously, and $R_{qq}$, the distance between points one quarter and three quarters along the nanotube from the end points used for $R_{ee}$. 
Specifically, $R_{qq}$ is defined as the distance between the centres of mass of another two groups of nucleotides, namely those nucleotides at positions 285--305 and 768--788, respectively. The nucleotide positions are as defined in the caDNAno design.

The sampling region chosen ranges from $R_{ee} = \SI{148.0}{nm}$ to $R_{ee} = \SI{190.7}{nm}$, and $R_{qq} = \SI{93.7}{nm}$ to $R_{qq} = \SI{139.7}{nm}$. Windows along $R_{ee}$ had a spacing of $\SI{0.658}{nm}$, which is the same as that used for the one-dimensional sampling, and windows along $R_{qq}$ had a spacing of $\SI{1.70}{nm}$. 

The final equilibrated configurations from the corresponding $R_{ee}$ windows in the 1D umbrella sampling simulations were used as the starting configuration for each window. The simulations for each window were constrained by a two-dimensional harmonic potential centred at the values of $R_{ee}$ and $R_{qq}$ corresponding to that window. The harmonic bias potential had stiffness $k_{ee}=\SI{17.12}{pN/nm}$ and $k_{qq}=\SI{1.71}{pN/nm}$ in the $R_{ee}$ and $R_{qq}$ coordinates, respectively. Each window was equilibrated for $5\times10^6$ steps before a production run of $5\times10^6$ steps. After the production run, two-dimensional WHAM \cite{Kumar95,wham} was performed to calculate the free-energy landscape in the transition region. 

\paragraph{Combining 1D and 2D umbrella sampling}
To combine the free-energy landscape from one- and two-dimensional umbrella sampling of the SST design, we also recorded $R_{qq}$ in the one-dimensional sampling runs, but without any bias on this order parameter. We performed two-dimensional WHAM on the whole data set, but excluding the data from the one-dimensional sampling that overlapped with that from the two-dimensional sampling. The two-dimensional free-energy landscape was then projected onto $R_{ee}$ to produce the new estimate of the one-dimensional landscape that is shown in Fig.\ \ref{fig:A(r)}.

\paragraph{Stretching}
We calculated the force-extension curve, $F(z)$, from our free-energy landscape, $A(R_{ee})$, using the following approach. Firstly, $F(z)$ can be related to $P(z)$ the probability distribution function for $z$:
\begin{flalign*}
    F(z) 
    &= -\frac{dA(z)}{dz} &\\
    &= -\frac{d(-k_{B}T\ln{P(z)})}{dz} &\\
    &= k_{B}T\frac{d\ln{P(z)}}{dz}  , &
\end{flalign*}  
where $A(z)$ is the free energy as a function of $z$. Secondly, $P(z)$ can be related to $P(r)$ the probability distribution function of $r$ 
\begin{flalign*}  
    &P(z) &\\
    &= \frac{dC(z)}{dz} &\\
    &= \frac{d}{dz} \left(1-\int_{0}^{2\pi}\int_{0}^{\arccos(\frac{z}{r})}\int_{z}^{\infty}P(r)r^2\sin{\theta} dr d\theta d\phi\right) &\\
    &= \frac{d}{dz}\int_{z}^{\infty}P(r) 2\pi r^2 \left(\frac{z}{r}-1\right) dr &\\
    &= \int_{z}^{\infty}P(r)2\pi r dr , &
\end{flalign*}  
where $C(z)$ denotes the cumulative distribution function for $z$. Finally, $P(r)$ can be related to $A(R_{ee})$:
\begin{flalign*}  
    P(r)
    &= \frac{1}{4\pi r^2}\exp\left({\frac{-A(R_{ee})}{k_{B}T}}\right). &
\end{flalign*}

\paragraph{Nanotube radius}
The average nanotube radius was calculated by averaging the radius at each slice (defined as the set of nucleotides with the same base-pair index in the caDNAno file) along the nanotube, where each slice is separated by a single base-pair step:
\[
\langle r_\mathrm{nanotube} \rangle = \frac{1}{N_{\text{slice}}}
\sum_{i=1}^{N_{\text{slice}}}r_{\mathrm{nanotube}}(i).
\]
The radius of the nanotube at a given slice is calculated by averaging the distances from the centre of each double helix to the centre of the nanotube for that slice: 
\[
r_{\mathrm{nanotube}}(i)=\frac{1}{N_{\mathrm{helix}}}\sum_{j=1}^{N_{\mathrm{helix}}} \left|\mathbf{R}_{\mathrm{duplex},j}(i)-\mathbf{R}_{\mathrm{nanotube}}(i)\right|, 
\]
where the nanotube centre is simply defined as
\[
\mathbf{R}_{\mathrm{nanotube}}(i)=\frac{1}{N_{\mathrm{helix}}} \sum_{j=1}^{N_{\mathrm{helix}}} \mathbf{R}_{\mathrm{duplex},j}(i).
\]
Following Ref.\ \citenum{Chhabra20} the centre of a double helix, $\mathbf{R}_{\mathrm{duplex}}$, is defined as
\begin{align*}
    \mathbf{R}_{\mathrm{duplex}}=
    &\frac{1}{2}\left(\mathbf{r}_{\mathrm{nuc1}}+\mathbf{r}_{\mathrm{nuc2}}\right)\\
    &+\frac{\alpha}{2}\left(\widehat{\mathbf{b}}_{\mathrm{nucl}} \times \widehat{\mathbf{n}}_{\mathrm{nuc} 1}+\widehat{\mathbf{b}}_{\mathrm{nuc} 2} \times \widehat{\mathbf{n}}_{\mathrm{nuc} 2}\right)
\end{align*}
where $\mathbf{r}$, $\widehat{\mathbf{b}}$, $\widehat{\mathbf{n}}$ are the centre of mass position, the base unit vector, and the normal unit vector of the oxDNA nucleotide (Fig.\ \ref{sfig:oxDNA2}), respectively, and $\alpha=0.06$ (in the oxDNA simulation unit of length). The second term is a correction factor that takes into account that the centre of mass of the two nucleotides in an oxDNA base pair is slightly displaced towards the minor groove.

\paragraph{Relaxation of a bent nanotube}
For all three designs, an equilibrated configuration from the last umbrella sampling window ($R_{ee} = 0$) was used as the starting point. With the biasing umbrella potential removed the configuration was free to relax back to equilibrium with $R_{ee}$ being used to monitor this relaxation. The simulation was run until the nanotube returned to its normal, relaxed length. 

\subsection{Origami sheets}
\begin{figure*}[t]
    \centering
    \includegraphics[width=0.97\textwidth]{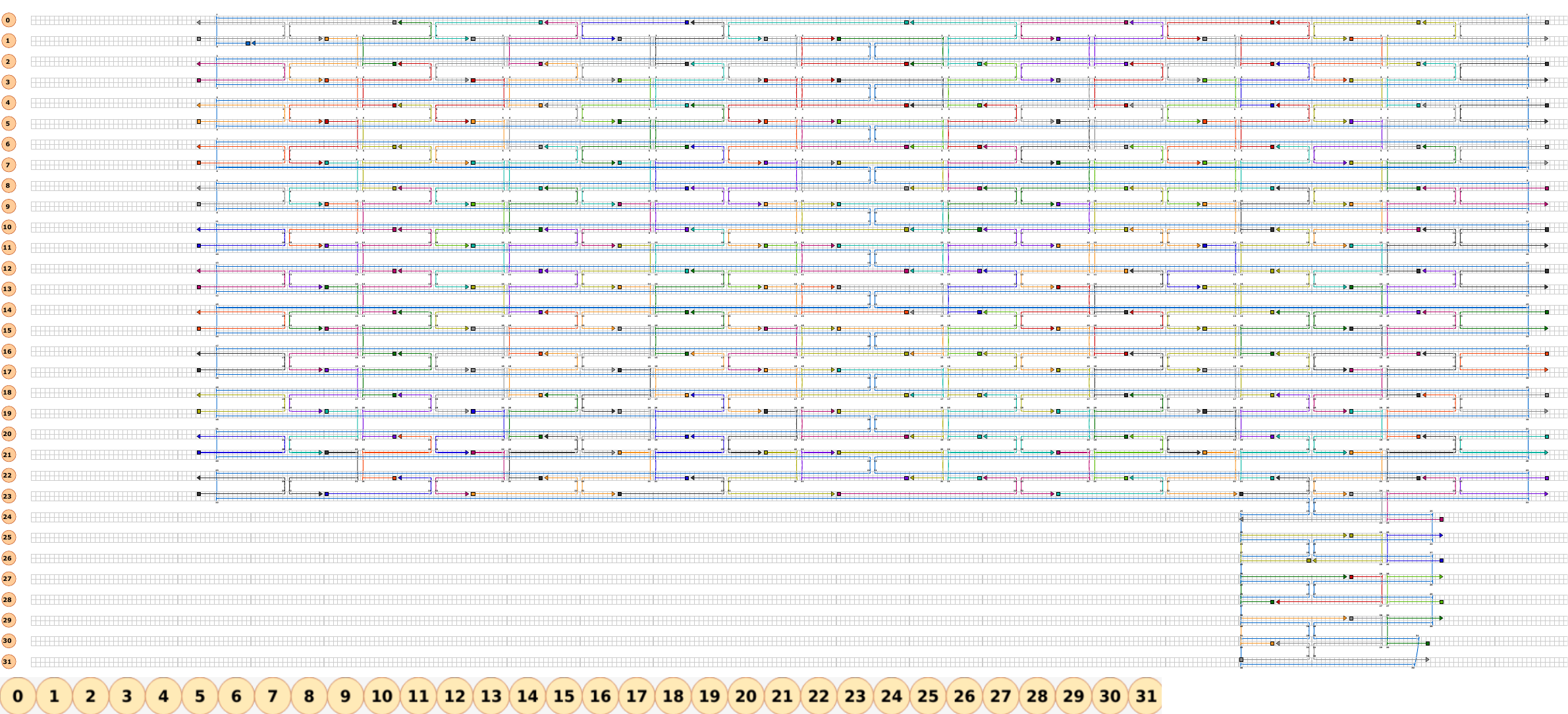}
    \includegraphics[width=0.97\textwidth]{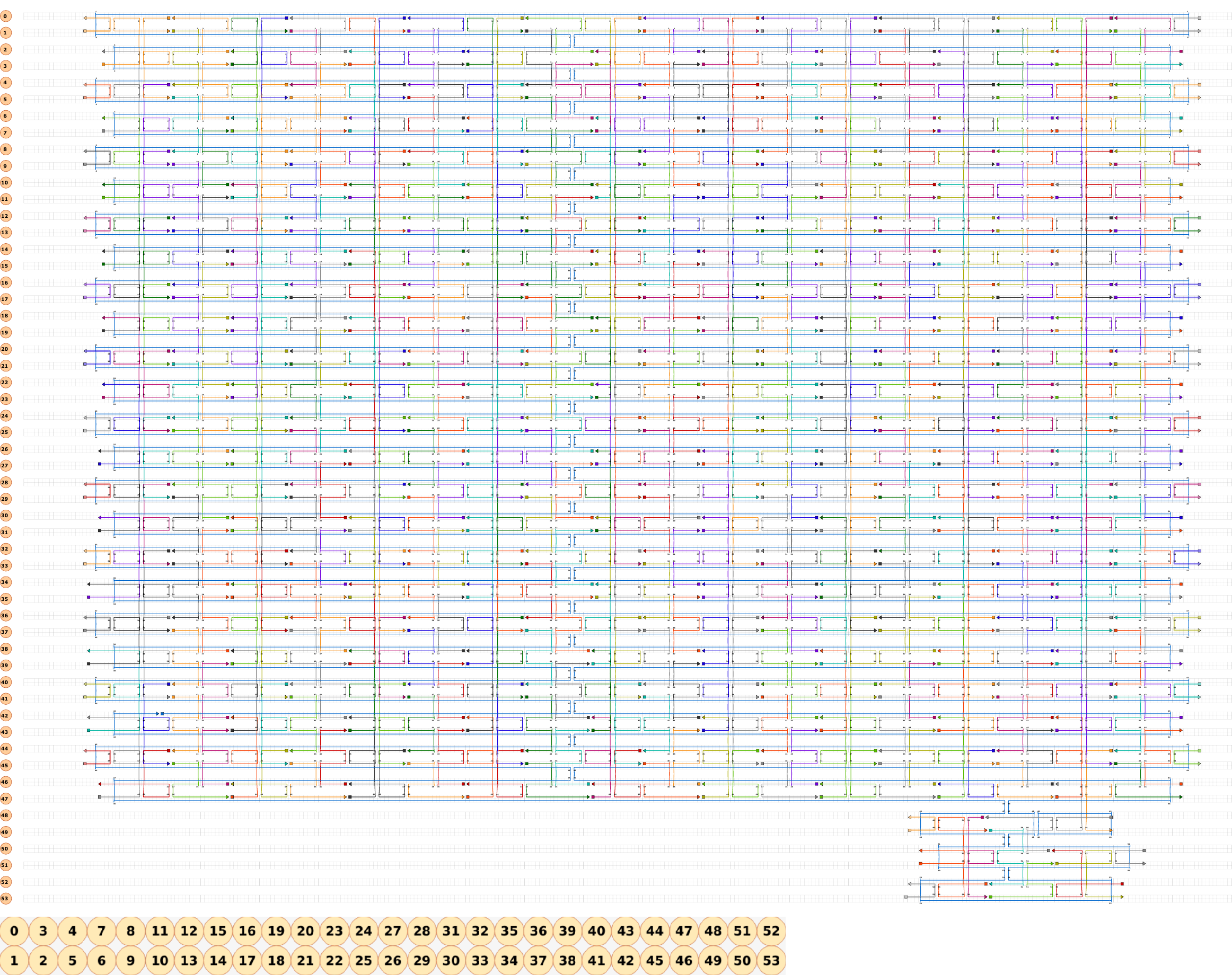}
    \caption{Full caDNAno designs for the single-layer and double-layer DNA origami sheets.}
    \label{sfig:cadnanoSI2}
\end{figure*}

\paragraph{2D Umbrella Sampling}
We used two-dimensional umbrella sampling to calculate the free-energy landscape as a function of the order parameters $R_1$ and $R_2$, the two diagonal distances of the sheet. To avoid effects associated with the greater splaying of the helices at the edges of the sheet and the larger fluctuations at the corners, we defined each diagonal distance as the distance between the centres of mass of two groups of nucleotides that are slightly in from the corners. For the single-layer sheet, $R_1$ is the distance between nucleotides in helices 2--3 and positions 72--103, and those in helices 20--21 and positions 264--295; $R_2$ is the distance between nucleotides in helices 20--21 and positions 72--103, and those in helices 2--3 and positions 264--295. For the double-layer sheet, $R_1$ is the distance between nucleotides in helices 4--7 and positions 48--63, and those in helices 42--45 and positions 256--271; $R_2$ is the distance between nucleotides in helices 42--45 and positions 48--63, and those in helices 4--7 and positions 256--271. The helices and nucleotide positions are as defined in the caDNAno design (Fig.\ \ref{sfig:cadnanoSI2}).

For each sheet design, we performed a set of simulations where the order parameters were restrained with a harmonic potential in each sampling window. The sampled ranges of the order parameter were from $R_1 = 0$ to $R_1 = R_{\mathrm{max}}$ and $R_2 = 0$ to $R_2 = R_{\mathrm{max}}$, where $R_{\mathrm{max}}$ is the diagonal distance when the sheet is flat.

Before the umbrella sampling simulations, we prepared starting configurations for each window using pulling simulations. The sheets were first stretched diagonally outwards until they were flat
to remove any initial bias towards the preferred curvature. Then, $R_1$ was restrained at $R_{\mathrm{max}}$ with a harmonic bias potential of stiffness $k=\SI{57.09}{pN/nm}$. The equilibrium position of the harmonic bias potential was gradually reduced at a constant rate such that it reaches zero after $10^8$ steps. $R_2$ was similarly restrained during the pulling, but at a constant value of $R_{\mathrm{max}}$. Configurations were outputted every $10^6$ steps, resulting in a total of 100 configurations. From each of these configurations, a similar pulling simulation was performed on $R_2$ while $R_1$ was restrained at a constant value corresponding to that configuration. These pulling simulations generated starting configurations for 10\,000 windows, each representing a point ($R_1,R_2$) on a square grid of 100 equally-spaced $R_1$ and $R_2$ values that run from 0 to $R_{\mathrm{max}}$.

After the starting configurations had been prepared, the simulations in each window were constrained by a 2-dimensional harmonic bias potential of stiffness $k=\SI{17.12}{pN/nm}$ centred at the value of $R_1$ and $R_2$ corresponding to that window. Each window was equilibrated for $10^6$ steps before a production run of $10^6$ steps. $R_1$ and $R_2$ was outputted every $10^3$ steps, giving $10^3$ data points for each window. The relatively short simulations were justified by the quick convergence of the data. For a smaller part of the landscape, we found that running the simulations for 10 times longer did not significantly alter the landscape.

Using the biased probability distributions of $R_1$ and $R_2$ in each window, we used 2D WHAM to calculate the unbiased free-energy landscape of the system. In total, 7500 windows were used for generating the free-energy landscape. Windows where both $R_1$ and $R_2$ are smaller than $R_\mathrm{max} / 2$ were discarded because the configurations were too strained and severely deformed.

\paragraph{Curvature analysis}
The surface of the origami sheet can be represented parametrically in terms of the helix index ($h$) and base pair index ($b$) in the caDNAno design. Mathematically, the surface is given by 
$\mathbf{r}(h,b) = ({X}(h,b), {Y}(h,b), {Z}(h,b))$, 
where ${X}$, ${Y}$ and ${Z}$ 
are the splines of the respective coordinates fitted as a function of $h$ and $b$. The principal curvatures, $\kappa_1$ and $\kappa_2$, are the minimum and maximum values of the curvature at a given point on the surface. The mean curvature is given by $H=(\kappa_1+\kappa_2)/2$ and the Gaussian curvature by $K=\kappa_1 \kappa_2$.

To calculate these quantities at a given point $(h,b)$ on the sheet, we use the formulae\cite{diff_geom}
\begin{align*}
H&=\cfrac{EN-2FM+GL}{2(EG-F^2)}  \\ 
K&=\frac{LN-M^2}{EG-F^2}
\end{align*}
where 
\begin{align*}
E &= \mathbf{r}_h\cdot\mathbf{r}_h, & 
F &= \mathbf{r}_h\cdot\mathbf{r}_b, \\
G &= \mathbf{r}_b\cdot\mathbf{r}_b, &
L &= \mathbf{r}_{hh}\cdot\widehat{\mathbf{n}}, \\
M &= \mathbf{r}_{hb}\cdot\widehat{\mathbf{n}}, & 
N &= \mathbf{r}_{bb}\cdot\widehat{\mathbf{n}}, 
\end{align*}
and
\begin{align*}
\mathbf{r}_h &= \cfrac{\partial\mathbf{r}}{\partial h}, & 
\mathbf{r}_b &= \cfrac{\partial\mathbf{r}}{\partial b}, \\
\mathbf{r}_{hh} &= \cfrac{\partial^2\mathbf{r}}{\partial h^2}, & 
\mathbf{r}_{hb} &= \cfrac{\partial^2\mathbf{r}}{\partial h\partial b}, \\
\mathbf{r}_{bb} &= \cfrac{\partial^2\mathbf{r}}{\partial b^2}, & 
\widehat{\mathbf{n}} &= \cfrac{\mathbf{r}_h \times \mathbf{r}_b}{|\mathbf{r}_h \times \mathbf{r}_b|}. 
\end{align*}
For each system, the curvature calculations were done on a configuration averaged over an unbiased simulation trajectory started from the relevant free-energy minimum.

\clearpage
\subsection{Further results}

\begin{figure*}[t]
    \centering
    \includegraphics[width=\textwidth]{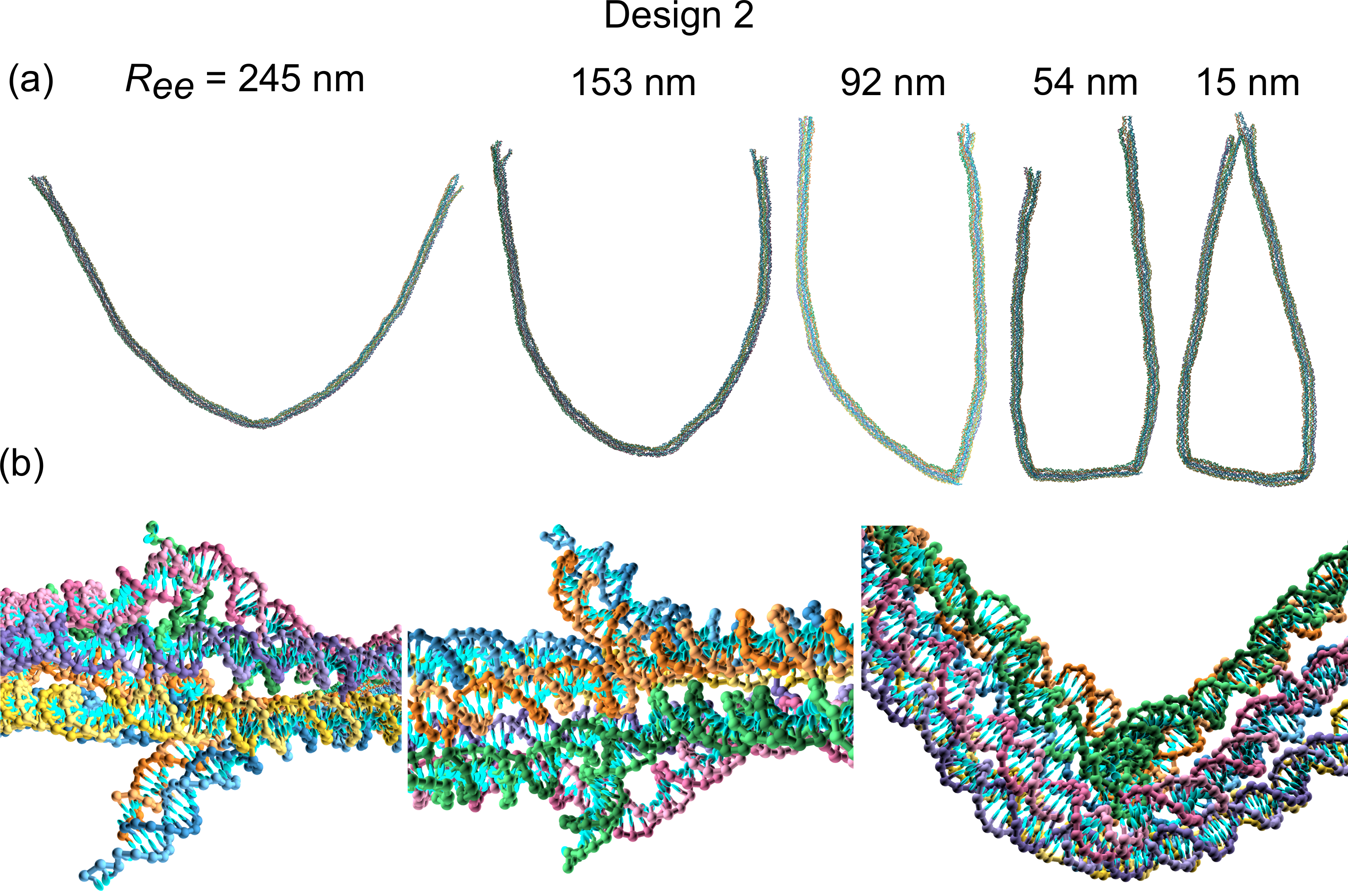}
    \caption{(a) Representative configurations of the design 2 nanotube at different values of $R_{ee}$. (b) Close-ups of the kink for the $R_{ee}=92$\,nm configuration. The view are from outside the kink (left), inside the kink (middle) and side-on (right).}
    \label{fig:design2_config}
\end{figure*}

\begin{figure*}
    \centering
    \includegraphics[width=\textwidth]{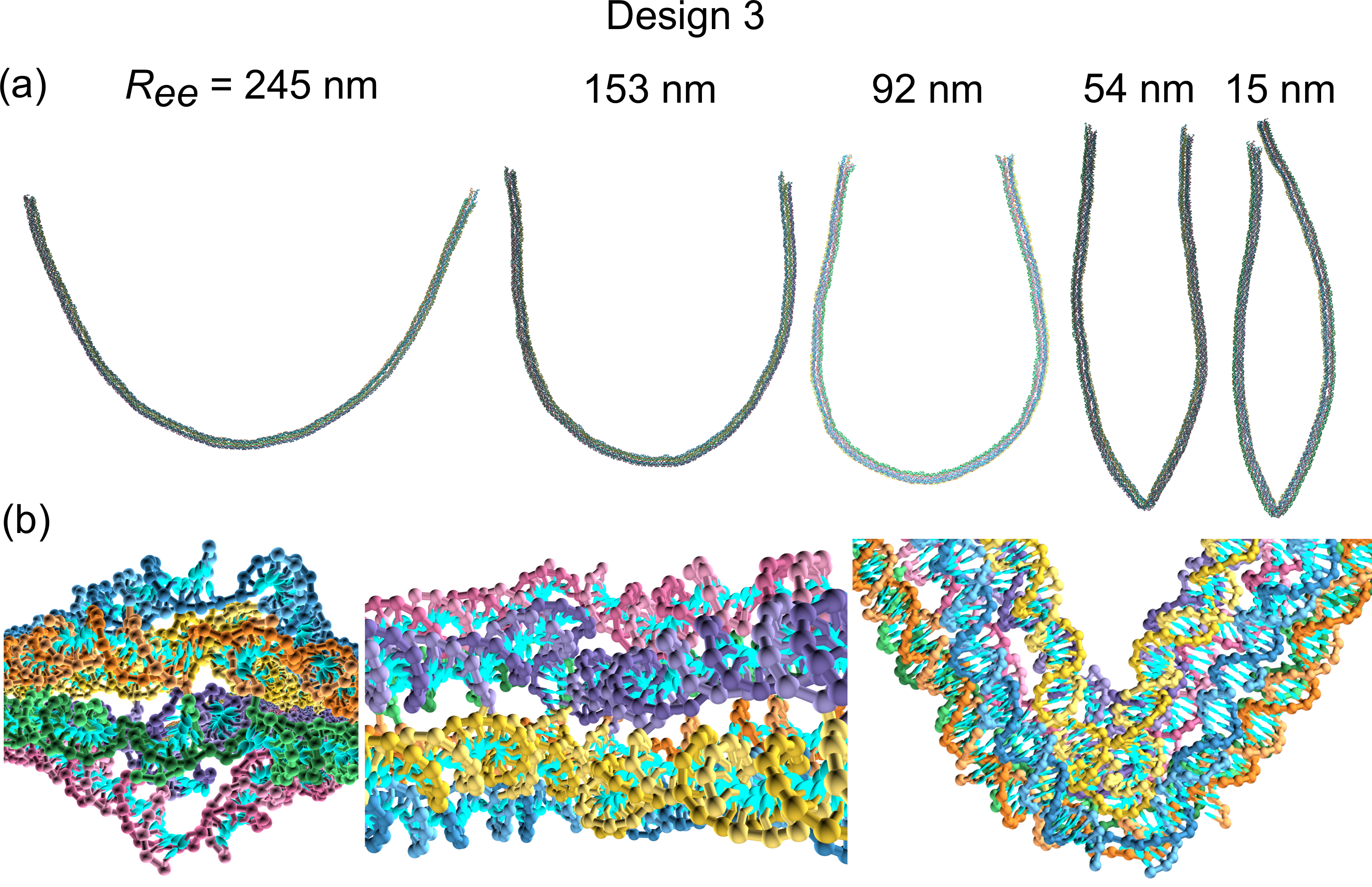}
    \caption{(a) Representative configurations of the design 3 nanotube at different values of $R_{ee}$. (b) Close-ups of the kink for the $R_{ee}=54$\,nm configuration. The view are from outside the kink (left), inside the kink (middle) and side-on (right).}
    \label{fig:design3_config}
\end{figure*}

\paragraph{Nanotube configurations}
Configurations of the design 1 nanotube at a number of values of $R_{ee}$ were shown in the main text (Fig.\ 1). Here, in Figs.\ \ref{fig:design2_config} and \ref{fig:design3_config} we similarly illustrate configurations of the design 2 and 3 nanotubes. Their reduced tendency to kink (compared to design 1) is apparent from the absence of kinks in the 153\,nm example for design 2 and the 153 and 92\,nm examples for design 3. The kinks in the design 2 nanotubes involve the unbinding of short 2-bp domains and perhaps also the end domain of a staple. By contrast, for the design 3 nanotubes the kinks involve the folding of the tube at the planes where each helix has a nick or a four-way junction, leading to the complete loss of the coaxial stacking at these sites.

\paragraph{Interacting nucleotides}
Three types of interactions are counted, namely hydrogen bonding, stacking, and coaxial stacking (Fig.\ \ref{sfig:oxDNA2}). A pair of nucleotides is considered as interacting through a particular interaction when the relevant interaction energy is lower than \SI{-2.49}{kJ/mol}. 

\begin{figure*}
    \centering
    \includegraphics[width=170mm]{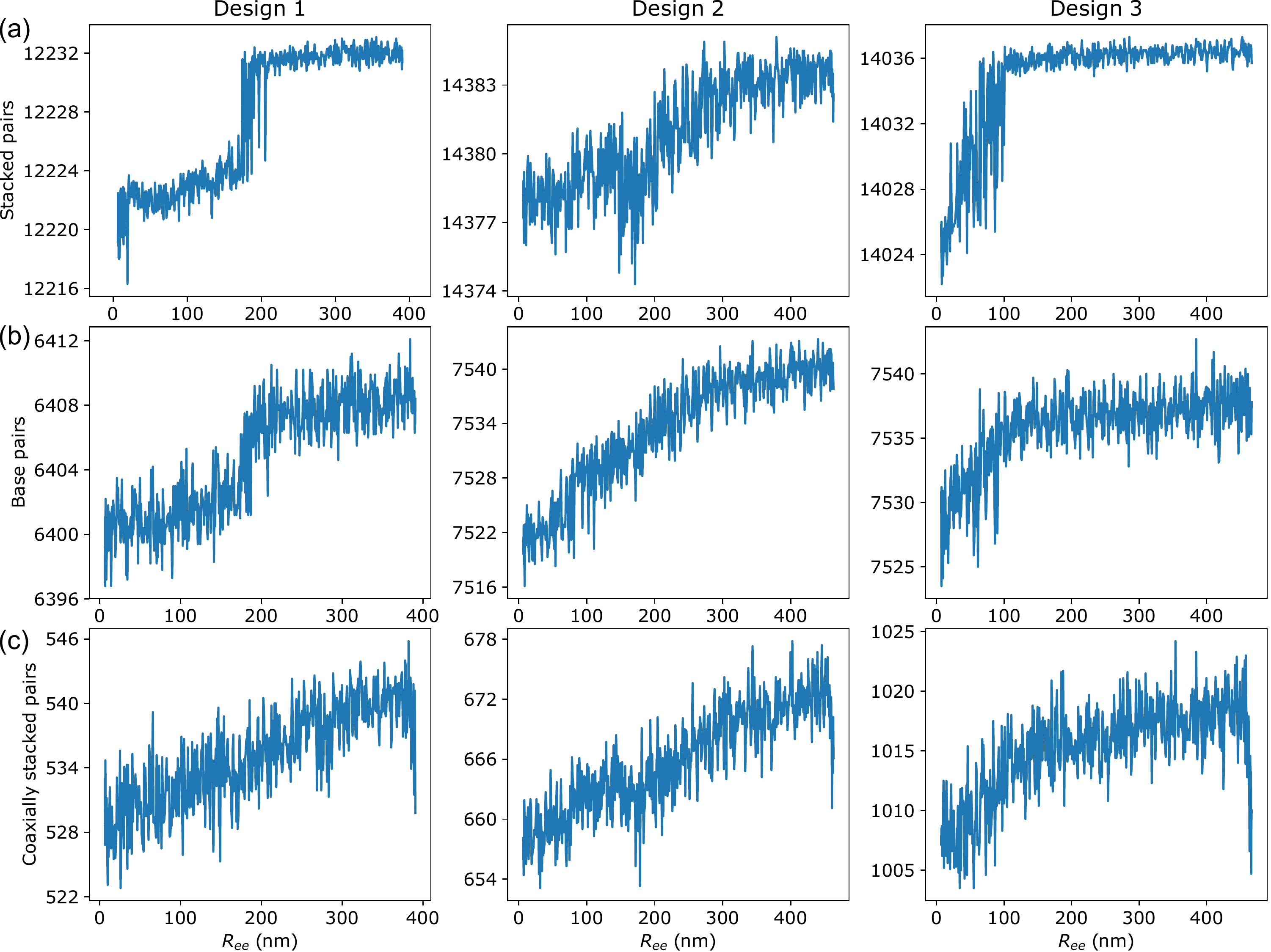}
    \caption{The total number of (a) pairs of stacked bases, (b) base pairs, and (c) coaxially-stacked pairs for the three DNA nanotube designs as a function of the end-to-end distance. Each point represents an average over one of the 500 umbrella sampling windows. }
    \label{sfig:interactions}
\end{figure*}

To produce the plots in Fig.\ \ref{sfig:interactions}, the total number of interacting pairs for each interaction type is counted for each configuration in the production trajectory in each umbrella sampling window, and averaged over each trajectory. 

\begin{figure*}
    \centering
    \includegraphics[width=\textwidth]{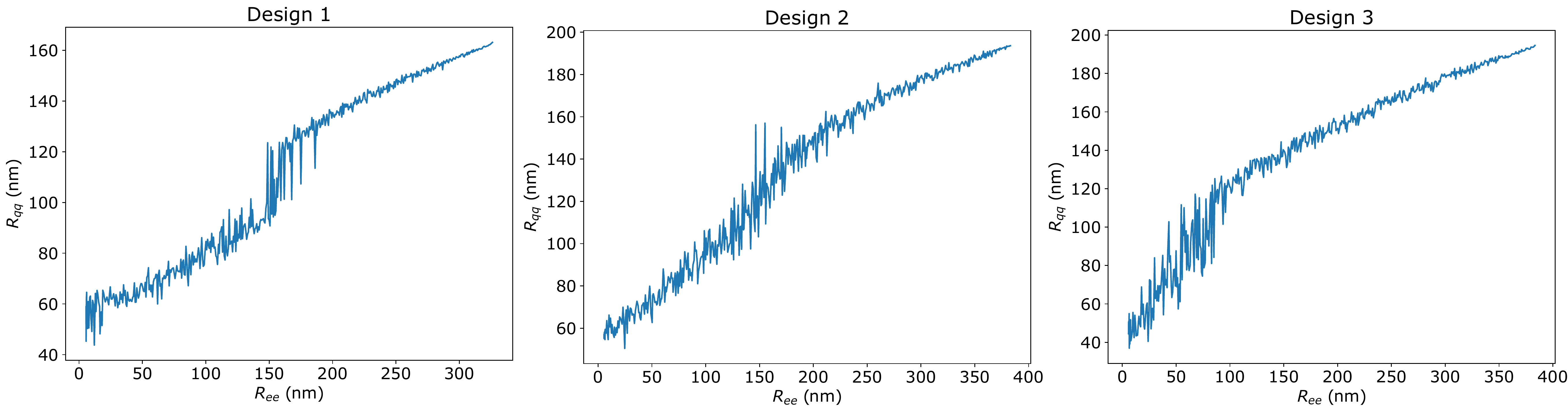}
    \caption{The \textonequarter--\textthreequarters\ 
    distance ($R_{qq}$) as a function of the end-to-end distance ($R_{ee}$) for the three DNA nanotube designs. Each point represents an average over one of the 500 umbrella sampling windows.}
    \label{sfig:r_quarter}
\end{figure*}

There are some common features in the plots of the $R_{ee}$-dependence of the number of interacting pairs for the three nanotube designs in Fig.\ \ref{sfig:interactions}. In the WLC regime of bending, the number of stacked pairs and base pairs remains fairly constant, while these numbers begin to decrease as a kink forms. Meanwhile, the number of coaxially stacked pairs decreases gradually even in the WLC regime. The numbers of stacked pairs and of base pairs show the clearest signals of kinking, with the most abrupt changes being observed for design 1. These can be correlated with the changes in $R_{qq}$ seen in the plots in Fig.\ \ref{sfig:r_quarter}. Comparing the three nanotube designs, design 1 starts to kink at the largest $R_{ee}$ and design 3 at the smallest $R_{ee}$.

For the stretched nanotubes, the number of stacked pairs and hydrogen bonds remains roughly constant, whereas the number of coaxially stacked pairs decreases sharply. This suggests that the junctions show greater compliance under tension than base pairs in the middle of a double helix. This effect may be coupled to the local structural changes at the junctions associated with the stretch-induced decrease in the nanotube radii. 

For nanotubes released from a bent configuration, after they have fully relaxed, the number of interacting pairs is essentially the same as that of a normal unstressed nanotube (Table \ref{table:interactions}). This suggests that strong bending and kink formation do not induce irreversible structural changes in the nanotubes.

\begin{table*}[ht!]
    \centering
        \begin{tabular}{ccccc}
            \hline\hline
            Design & State & Stacked pairs & Base pairs & Coaxially stacked pairs \\ \hline
             & at equilibrium & 12233 & 6409 & 540 \\ 
            1 & bent and released & 12232 & 6408 & 540\\ 
             & theoretical maximum & 12234 & 6426 & 606 \\ \hline
             & at equilibrium & 14383 & 7540 & 672 \\
            2 &bent and released & 14384 & 7538 & 673\\ 
             & theoretical maximum & 14386 & 7560 & 722 \\ \hline
             & at equilibrium & 14036 & 7536 & 1020 \\ 
            3 & bent and released & 14036 & 7536 & 1019\\ 
             & theoretical maximum & 14038 & 7560 & 1070 \\ 
            \hline\hline
        \end{tabular}
    \caption{The average number of interaction pairs of different types for the three DNA nanotube designs. These are given for the relaxed nanotube at equilibrium and a nanotube that has been allowed to relax back towards equilibrium from an initially highly bent configuration in an unbiased simulation (i.e.\ the end configurations for the simulations represented in Fig.\ 6). In addition, the theoretical maximum values allowed by the design have been provided. }
    \label{table:interactions}
\end{table*}

\begin{figure*}
    \centering
    \includegraphics[width=\textwidth]{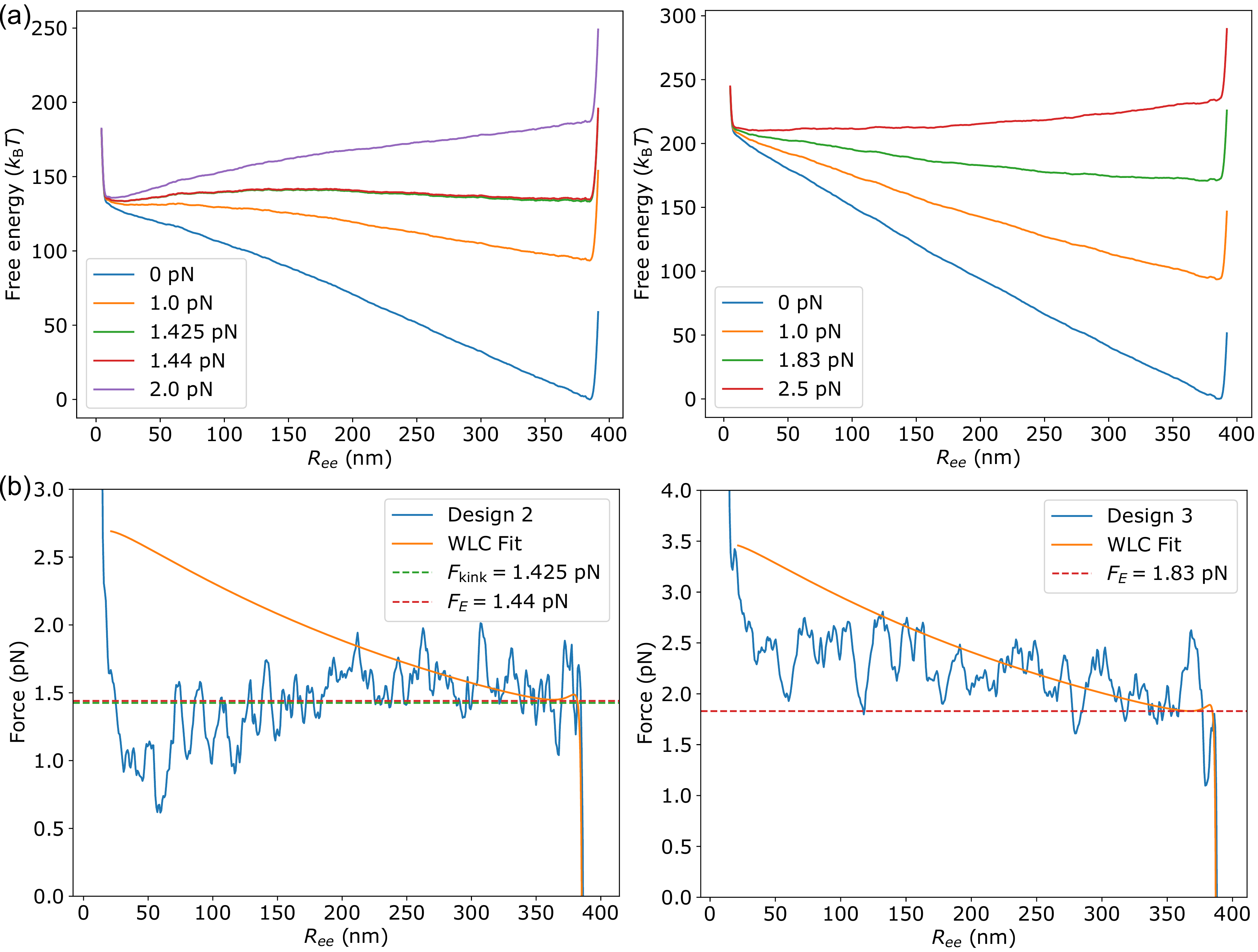}
    \caption{(a) The free-energy landscape for the design 2 and design 3 nanotubes at a series of compressive forces along the end-to-end vector.
    (b) $F(R_{ee})=-dA(R_{ee})/dR_{ee}$ for the computed landscapes and their WLC fits.
    The horizontal lines correspond to the force at which the two free-energy minima associated with the kinked and unkinked states are degenerate (design 2 only) and the predicted Euler buckling critical force. 
    }
    \label{sfig:force_2_3}
\end{figure*}

\paragraph{Force-dependent landscapes}
In the main text we showed how the free-energy landscape of nanotube 1 changes as a compressive force is applied along the end-to-end vector (Fig.\ 3). Here, we show the equivalent plots for the design 2 and 3 nanotubes (Fig.\ \ref{sfig:force_2_3}). The behaviour in the homogeneously-bent regime is very similar to the design 1 nanotube. Namely, the free-energy landscape becomes very flat at the Euler critical buckling force, thus leading to a rapid change in $R_{ee}$ expected for the homogeneously-bent state as the force goes above $F_E$. The precise values of the critical buckling forces are different for all three systems due to their differences in persistence and contour lengths, as predicted by the formula given in the main text.

The main qualitative differences in behaviour between the three systems are due to their different propensities to kink. For the design 1 nanotube, the kinked state becomes lower in free energy than the unkinked state well before the critical buckling force is reached (Fig.\ 3). For the design 2 nanotube there is still a force range for which the kinked and unkinked states are both free-energy minima, but the force at which they are degenerate is now very close to $F_E$ (at $F_\mathrm{kink}$ the barrier between the two states is $8.0\,k_BT$). Finally, for design 3 the kinked state only becomes stable above $F_E$ and at no point does the landscape have two free-energy minima separated by significant barrier. These changes are in part simply due to the larger contour length of these two nanotubes, which leads to a lowering of $F_E$, but the slope of the (zero-force) free-energy landscape in the kinked regime, i.e.\ the free-energy cost of further bending a kink, is also important as this determines the force at which the kinked state becomes a minimum. 

\begin{figure*}
    \centering
    \includegraphics[width=\textwidth]{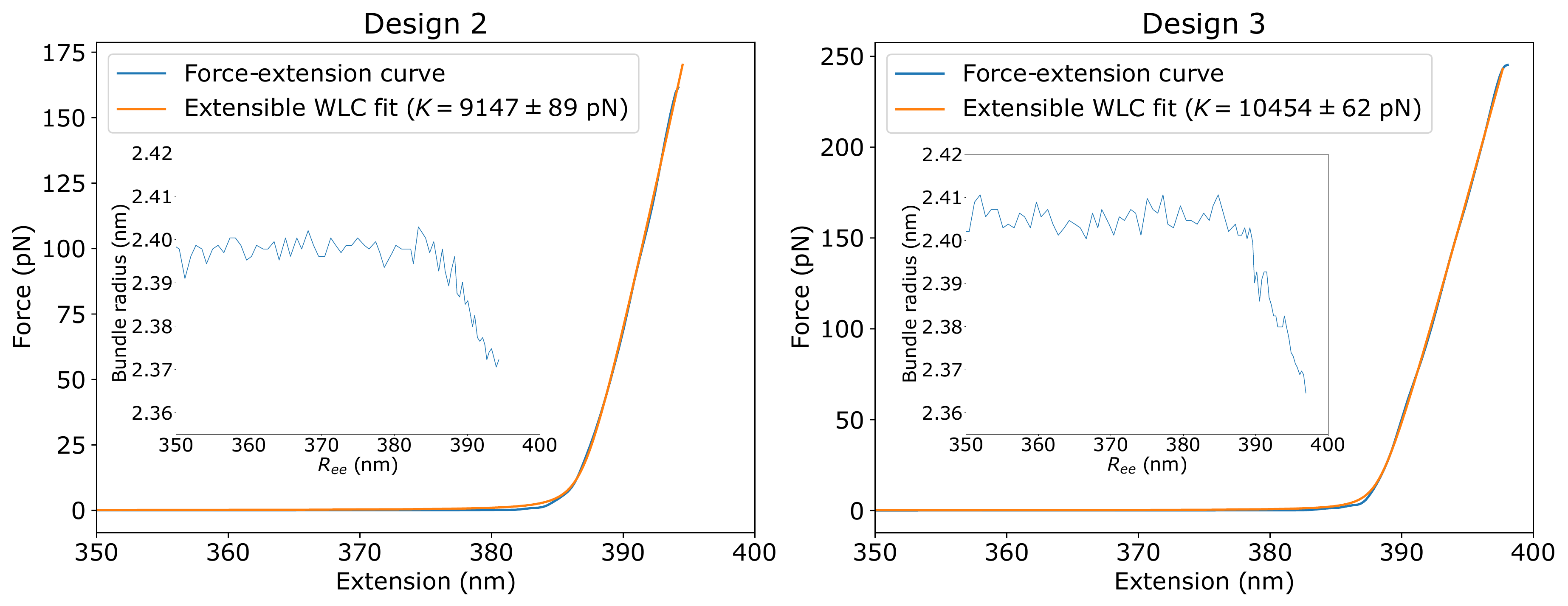}
    \caption{Force-extension curves for the DNA nanotube designs 2 and 3 along with extensible WLC fits (the fit parameters are given in Table \ref{table:ewlc_fit}). The insets show the average nanotube radius as a function of $R_{ee}$.}
    \label{sfig:F(z)_supp}
\end{figure*}

\paragraph{DNA nanotube stretching}
In the main text we showed the force-extension curve for the design 1 nanotube as well as the change in radius of the nanotube on stretching (Fig.\ 6). Here, we provide similar plots for the design 2 and 3 nanotubes (Fig.\ \ref{sfig:F(z)_supp}). The behaviour is again very similar. The extensible worm-like chain model provides an excellent fit to the force-extension curves, and both nanotubes again show clear decreases in their radii in response to their stretching.

\paragraph{WLC fits}
The parameters for the WLC fits to $p(R_{ee})$ using the form given in Ref.\ \citenum{Becker10} are given in Table \ref{table:wlc_fit} for the three nanotube designs. The parameters for the extensible WLC fits to the force-extension curves using the form given in Ref.\ \citenum{Odijk95} are given in Table \ref{table:ewlc_fit} for the three nanotube designs. The values obtained for the contour lengths from the two methods are consistent albeit with a much greater degree of precision available from the force-extension curves. Similarly, the values obtained for the persistence length are consistent. In this case, because the persistence lengths are much larger than the contour length, the force-extension curves provide much lower precision estimates of the persistence length.

For six DNA double helices stretched in parallel the extensional modulus would be expected to be six times the stretch modulus of the individual helices,\cite{Chhabra20} i.e.\ $6\times 2700\,\mathrm{pN}\approx 16\,000\,\mathrm{pN}$. However, due mainly to the change in the geometry of the crossovers linking the helices on stretching the extensional moduli of the nanotubes are instead about 10\,000\,pN. The loss of about 1\% of the coaxial stacking interactions on stretching (Fig.\ \ref{sfig:interactions}) may also play a small role in this reduction.

\begin{table*}
    \centering
        \begin{tabular}{cccc}
            \hline\hline
            & $L_c$ (nm) & $L_p$ (\SI{}{\micro\metre}) \\ 
            \hline
            Design 1 & $325 \pm 5$ & $6.49 \pm 0.14$ \\ 
            Design 2 & $388 \pm 6$ & $5.30 \pm 0.13$ \\ 
            Design 3 & $389 \pm 9$ & $6.77 \pm 0.24$ \\ 
            \hline\hline
        \end{tabular}
    \caption{Parameters for the WLC fit to probability distribution for the end-to-end distance for the three DNA nanotube designs. $L_c$ is the contour length.}
    \label{table:wlc_fit}
\end{table*}

\begin{table*}
    \centering
        \begin{tabular}{cccc}
            \hline\hline
            & $K$ (pN) & $L_c$ (nm) & $L_p$ (\SI{}{\micro\metre}) \\ 
            \hline
            Design 1 & $10077 \pm 150$ & $325.20 \pm 0.08$ & $5.63 \pm 0.48$ \\ 
            Design 2 & $9147 \pm 89$ & $387.73 \pm 0.07$ & $5.00 \pm 0.46$ \\ 
            Design 3 & $10454 \pm 62$ & $388.93 \pm 0.06$ & $6.07 \pm 0.74$ \\ 
            \hline\hline
        \end{tabular}
    \caption{Parameters for the extensible WLC fit to the force-extension curves of the three DNA nanotube designs.}
    \label{table:ewlc_fit}
\end{table*}

\begin{figure*}
    \centering
    \includegraphics[width=\textwidth]{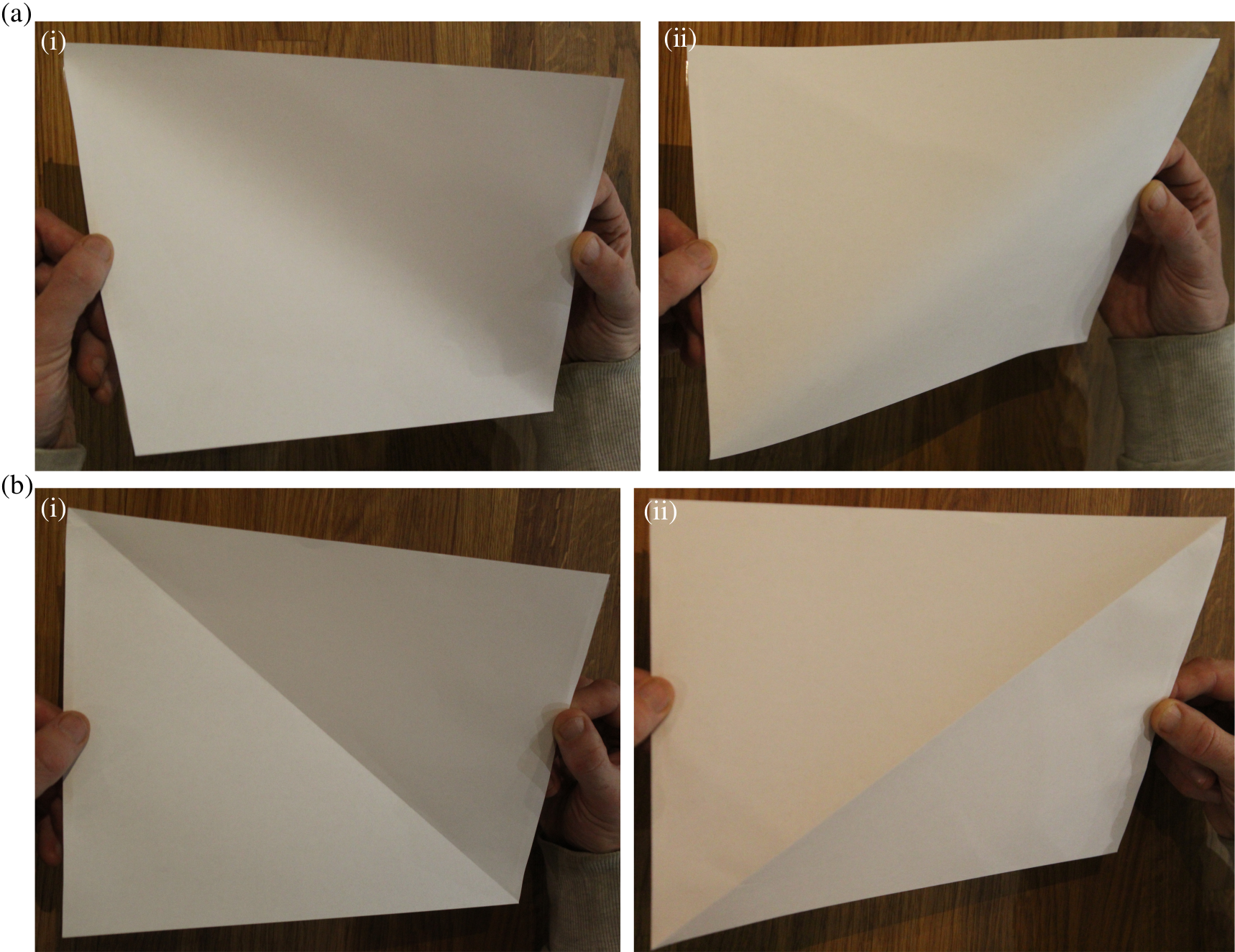}
    \caption{Images of right-twisted A4 sheets of paper. In (a) the curvature is continuous, whereas in (b) the curvature is localized at a fold. In (i) the bending is around the top-left to bottom-right diagonal and the curvature is positive, whereas in (ii) the bending is around the bottom-left to top-right diagonal and the curvature is negative. (b)(i) corresponds to a valley fold and (b)(ii) to a mountain fold. }
    \label{sfig:twistedA4}
\end{figure*}

\paragraph{Gaussian curvature of sheets}

\begin{figure*}
    \centering
    \includegraphics[width=\textwidth]{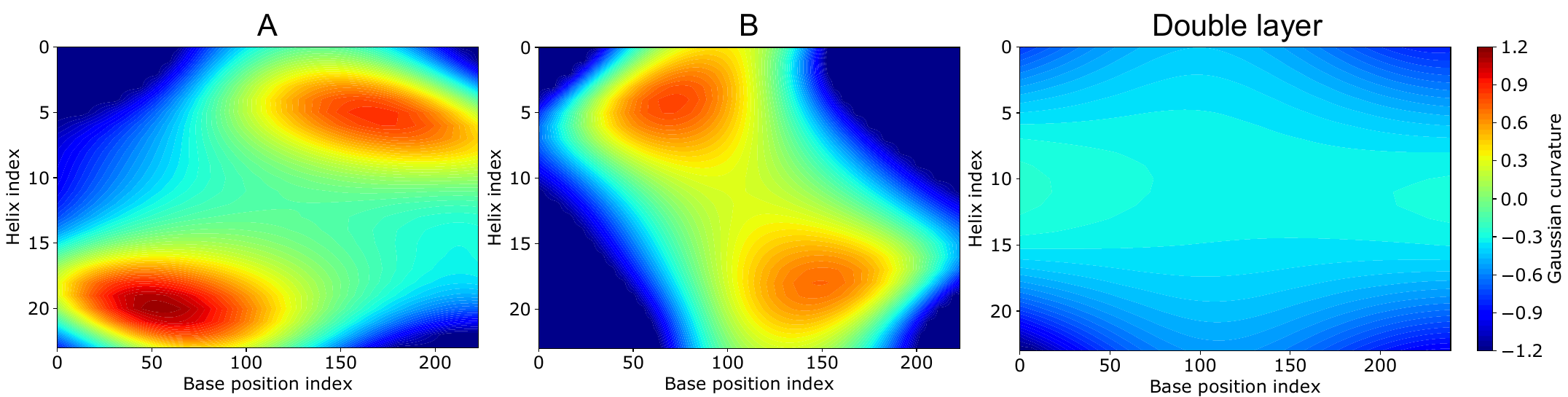}
    \caption{The Gaussian curvature of configurations at the two free-energy minima of the single-layer sheet and the free-energy minimum of the double-layer sheet as a function of the base-pair position within the sheet.}
    \label{sfig:gaussian_curvature}
\end{figure*}

In Fig.\ \ref{sfig:gaussian_curvature} we show the Gaussian curvature, which is the product of the principal curvatures, for the origami sheets. For the double-layer sheet, as expected from its saddle-like shape, it has a negative Gaussian curvature across the whole sheet; the Gaussian curvature is fairly constant in the centre of the sheet, but increases in magnitude towards the corners. For the two free-energy minima of the single-layer sheet, like for the mean curvature shown in Fig.\ 7(d) in the main text, there is an approximate symmetry with respect to the ``unperturbed'' diagonal. For both forms, there are zones of strong positive Gaussian curvature (the principle curvatures have the same sign) that correspond to the areas with greatest mean curvature. Along the unperturbed diagonal, the curvature is close to zero in the centre leading to a near-zero Gaussian curvature, but at the corners of this diagonal the curvature along the diagonal becomes opposite to the prevailing curvature, leading to strongly negative Gaussian curvature. The latter is evident in the configurations visualized in Fig.\ 7(c).

\end{document}